\documentclass[fleqn,usenatbib]{mnras}

\usepackage{newtxtext,newtxmath}

\usepackage[T1]{fontenc}

\DeclareRobustCommand{\VAN}[3]{#2}
\let\VANthebibliography\thebibliography
\def\thebibliography{\DeclareRobustCommand{\VAN}[3]{##3}\VANthebibliography}

\usepackage{graphicx}	
\usepackage{amsmath}	
\usepackage{amssymb}	
\usepackage{hyperref}

\newcommand{\xmm}{{\it XMM-Newton\/}~}
\newcommand{\chan}{{\it Chandra\/}~}
\newcommand{\swift}{{\it Swift\/}~}
\newcommand{\hst}{{\it HST\/}~}
\newcommand{\pgcsrc}{4XMM J091948.8-121429~}
\newcommand{\ngcsrc}{4XMM J112054.3+531040~}
\newcommand{\cv}{4XMM J165251.5-591503~}
\newcommand{\ctps}{\rm ~ct~s^{-1}}
\newcommand{\ergcms}{\rm ~erg~cm^{-2}~s^{-1}}
\newcommand{\ergs}{\rm ~erg~s^{-1}}
\newcommand{\pcmsq}{\rm ~cm^{-2}}

\title[Characterising three new eULXs]{Digging a little deeper: characterising three new extreme ULX candidates}

\author[T. P. Roberts et al.]{
T.\,P. Roberts$^{1}$\thanks{E-mail: t.p.roberts@durham.ac.uk},
D.\,J. Walton$^{2}$,
A.\,D.\,A.\, Mackenzie$^{1}$,
M. Heida$^{3}$,
S. Scaringi$^{1}$
\\
$^{1}$Centre for Extragalactic Astronomy \& Dept of Physics, Durham University, South Road, Durham DH1 3LE, UK\\
$^{2}$Centre for Astrophysics Research, University of Hertfordshire, College Lane, Hatfield AL10 9AB, UK\\
$^{3}$European Southern Observatory, Karl-Schwarzschild-Str 2, 85748 Garching bei M\"{u}nchen, Germany
}

\date{Accepted 2023 July 31. Received 2023 July 31; in original form 2023 June 2}

\pubyear{2023}

\begin{document}
\label{firstpage}
\pagerange{\pageref{firstpage}--\pageref{lastpage}}
\maketitle

\begin{abstract}
A prime motivation for compiling catalogues of any celestial X-ray source is to increase our numbers of rare sub-classes.  In this work we take a recent multi-mission catalogue of ultraluminous X-ray sources (ULXs) and look for hitherto poorly-studied ULX candidates that are luminous ($L_{\rm X} \geq 10^{40} \ergs$), bright ($f_{\rm X} \geq 5 \times 10^{-13} \ergcms$) and have archival \xmm data.  We speculate that this luminosity regime may be ideal for identifying new pulsating ULXs (PULXs), given that the majority of known PULXs reach similar high luminosities.  We find three sources that match our criteria, and study them using archival data.  We find \cv to possess a bright and variable Galactic optical/IR counterpart, and so conclude it is very likely to be a foreground interloper.  \pgcsrc does appear an excellent ULX candidate associated with the dwarf irregular galaxy PGC 26378, but has only one detection to date with low data quality.  The best dataset belongs to \ngcsrc which we find to be a moderately variable, spectrally hard ($\Gamma \approx 1.4$) X-ray source located in a spiral arm of NGC 3631.  Its spectral hardness is similar to known PULXs, but no pulsations are detected by accelerated pulsation searches in the available data.  We discuss whether other missions provide objects for similar studies, and compare this method to others suggested for identifying good PULX candidates.  
\end{abstract}




\section{Introduction}

Ultraluminous X-ray sources (ULXs; see \citealt{kfr17}) remain a compelling class of sources to study, despite more than two decades elapsing since the realisation that new and exciting astrophysics were required to explain their extraordinary X-ray luminosities of more than $10^{39} \rm ~erg~s^{-1}$ (e.g. \citealt{king01}).  In fact, our understanding has evolved substantially from the original focus on ULXs as intermediate-mass black hole candidates (IMBHs; \citealt{colbert99}) to the detection of pulsations in some objects (Pulsating ULXs, or PULXs, e.g. \citealt{bachetti14}).  The pulsations reveal the presence of a neutron star (NS) and hence very extreme accretion rates, that can reach apparent factors $\sim 500$ above Eddington \citep{israel17a,fuerst17}.  The observed phenomenology of ULXs is also suggestive of super-Eddington accretion, with the observed X-ray spectra \citep{gladstone09,bachetti13}, the correlated spectral and temporal variability behaviour \citep{sutton13}, the detection of fast outflows \citep{pinto16} and the presence of large bubble nebulae \citep{pakull02} strongly supporting this interpretation; this makes ULXs important to study in the context of the rapid formation of the earliest supermassive black holes (e.g. \citealt{banados18}).  ULXs may also have links with other exotic phenomena, for example they may constitute an evolutionary phase in the binary stellar systems that ultimately merge and are detected as gravitational wave sources \citep{mondal20a} and may be the systems responsible for Fast Radio Bursts \citep{sridhar21}.

The requirement for novel accretion physics is particularly pertinent for explaining ULXs that appear with luminosities in the $10^{40} - 10^{41} ~\rm erg~s^{-1}$ regime, sometimes described as {\it extreme ULXs\/}, or eULXs \citep{gladstone13}.   This is a distinct, separate class from the yet more luminous and rarer hyperluminous X-ray sources (HLXs) that appear above $10^{41} ~\rm erg~s^{-1}$ and remain the best candidates to host IMBHs, the archetype being ESO 243-49 HLX-1 \citep{farrell09}.  Given their extreme luminosities, eULXs are relatively well-studied and provide many of the archetypes for ULX behaviour that drive our understanding of it (e.g. Ho IX X-1, Ho II X-1, NGC 5408 X-1, NGC 1313 X-1 etc; see \citealt{kfr17} and references therein).  Data from such objects have driven many of the breakthroughs in understanding ULXs, from the curvature in their spectra above 2 keV (e.g. Ho IX X-1, \citealt{gladstone09}), to the detection of absorption lines from outflowing gas travelling at $v \approx 0.2c$ (e.g. NGC 1313 X-1, \citealt{pinto16}) and the detection of pulsations indicative of NSs (e.g. M82 X-2, \citealt{bachetti14}).  Indeed, the majority of known PULXs reach this regime at their brightest, these being M82 X-2; NGC 1313 X-2 \citep{sathyaprakash19}; NGC 7793 P13 \citep{fuerst16}; and NGC 5907 ULX, which peaks in the HLX regime \citep{israel17a}.  These luminosities are indicative of NSs with accretion rates at least $50-100$ times their Eddington limit.

Many of the current key questions for ULXs relate to PULXs.  The demographics of ULXs remain uncertain in terms of the proportion of the overall population that host a NS rather than a BH, with some suggestions that NS may dominate (e.g., \citealt{pintore17,koliopanos17,middleton17,walton18}).  The astrophysics of super-Eddington accretion onto NSs also remains a matter for debate, with the magnetic field strength and configuration (dipole/quadrupole) and the extent to which classical super-critical accretion models apply, in which a geometrically thick inner disc forms and a massive radiation-pressure driven wind is ejected from its surface, all key areas of uncertainty (e.g., \citealt{kluzniak15,mushtukov15,dall'osso15,mushtukov19}).  A key to progressing all these issues will be to find more PULXs, that will add to our population statistics and provide new observations that will help constrain our models of accretion on to neutron stars.  In this paper we examine whether the observational quirk that most detected PULXs reach luminosities above $10^{40} \ergs$ can be leveraged to find new PULXs, basing our search around the recent large multi-mission ULX catalogue of \cite{walton22} (hereafter W22).  The paper is arranged as follows.  We discuss the selection of targets in Section~\ref{sec:source_select}, and the reduction of the X-ray data for our chosen targets in Section~\ref{sec:reduction}.  The results are laid out in Section~\ref{sec:results}, and discussed in Section~\ref{sec:discussion}, before the paper is concluded.

\section{Source selection}
\label{sec:source_select}

\begin{figure*}
	\includegraphics[width=12cm]{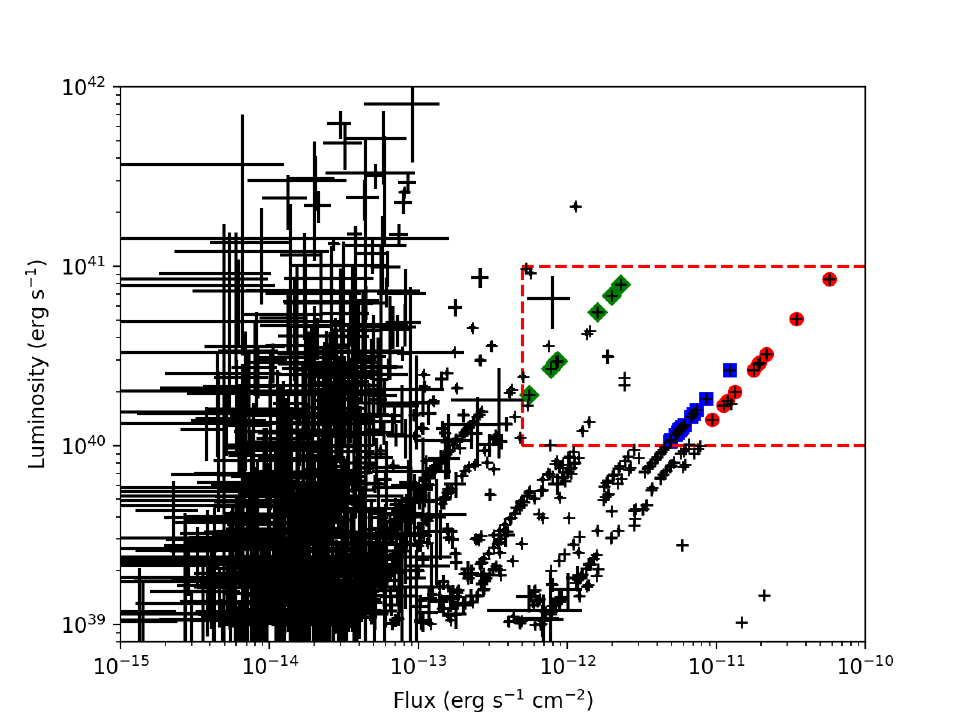}
    \caption{The selection of interesting sources.  The high flux, high luminosity region used to select sources is delineated by a dashed red line.  Source positions in this parameter space are marked by a black cross, with size matched to the $1\sigma$ uncertainties on these quantities (with a minimum cross size adopted for display purposes for those objects with extremely well-constrained values).  Three sources are highlighted for appearing multiple times in the region of interest: they are: red circles -- M82 X-1; blue squares -- NGC 1313 X-1; green diamonds -- NGC 5907 ULX.  The diagonal tracks are a trivial result of the scaling of flux by distance squared to give luminosity.}
    \label{fig:source_select}
\end{figure*}

The selection of interesting targets began with our recent multi-mission ULX catalogue, W22, and a simple observation: that 4 out of 6 known extragalactic PULXs have been seen to exceed a luminosity of $10^{40} \rm ~erg~s^{-1}$.  This compares to 1 in 4 ULX detections exceeding this threshold from 4XMM-DR10 and 2SXPS, and 1 in 6 from CSC2.0, in the W22 catalogue.  Even more starkly, an integration in the ULX regime, both above and below $10^{40} \rm ~erg~s^{-1}$, of the X-ray luminosity function of X-ray binaries in star forming galaxies from \cite{mineo12} shows that only 1 in 9 ULXs should appear above this threshold luminosity.  However, a bias to high luminosity ULXs in pulsation detections should perhaps be expected, given those detections require photon-rich data, and that is more immediately forthcoming from higher luminosity sources at a given distance and exposure length.  Nevertheless, this preponderance presents an interesting basis from which to search for new PULXs.  We therefore filtered W22 to select ULXs with both a luminosity between $10^{40}$ and $10^{41} \rm ~erg~s^{-1}$, i.e. in the eULX range, and an observed flux sufficiently high that a pulsation detection could be possible given a moderately deep observation, which we set at $5 \times 10^{-13} \rm ~erg~cm^{-2}~s^{-1}$\footnote{PULX detections tend to come from data with $\gtrsim 10000$ pn counts (G. Israel, priv. comm.); at a minimum flux of $5 \times 10^{-13} \rm ~erg~cm^{-2}~s^{-1}$ and a hard, PULX-like spectrum (e.g. power-law continuum with $\Gamma \sim 1.5$ and column $N_{\rm H} \sim 2 \times 10^{21} \pcmsq$) this can be obtained in $\sim 100$ ks, i.e. within a single \xmm orbit, even if we allow for some data loss to background flares (see Section~\ref{sec:reduction}).  We note that such datasets are rare in the archives for all but a handful of well-studied ULXs; here we focus on analysing the currently available data, with a view to proposing for the necessary observations if any interesting sources are identified.}.  We also limit our analysis to \xmm data, given that the pn is the only detector currently operating in the 0.5-10\,keV band with the combination of effective area and readout time that regularly permits the detection of ULX pulsations (we discuss what similar selection criteria reveal for \swift and \chan in Section~\ref{sec:discussion}).

We show the flux-luminosity parameter space for the {\it XMM-Newton} detections in W22 in Fig.~\ref{fig:source_select}, with the region of interest delineated.  The fluxes used are taken directly from the 4XMM-DR10 EP\_8\_FLUX column, which is the flux in the broad 0.2-12\,keV band (cf. \citealt{webb20}) , with the luminosities calculated directly from these fluxes using the distances assumed in W22.  Note there are many other ULX detections at similar fluxes but lower luminosities, which will be the subject of future work.  There is also one detection at high flux in the HLX regime; this object will be included in a new study of HLXs (Mackenzie et al., in prep.).  In total, 49 detections of of 15 ULXs are within the interesting parameter space; we list these objects in Table~\ref{tab:srclist}.  Three objects dominate the detections by number, and are highlighted in the figure: M82 X-1 (10 detections), NGC 1313 X-1 (16) and NGC 5907 ULX1 (7).  These are amongst the best-studied ULXs (for example references, see Table~\ref{tab:srclist}), and indeed the majority of these objects (12/15) are already well-studied in the literature.  However, three of the eULXs -- 4XMM J091948.8-121429, 4XMM J112054.3+531040 and 4XMM J165251.5-591503 -- have not been examined in detail before (with 4XMM J112054.3+531040 the only to have been previously-catalogued).  These three sources are the subject of the remainder of this work.

\begin{table*}
\centering
\caption{ULXs detected with high flux and high luminosity by \xmm in W22}
\label{tab:srclist}
\begin{tabular}{lcccccc}
\hline
Source ID	& Host galaxy	& $d$ $^a$	& $L_{\rm X,peak}$ $^b$	& \# det\underline{n}s $^c$	& Other name $^d$	& References $^d$ \\
 (4XMM J...)	& & (Mpc)	& ($\times 10^{40} \rm ~erg~s^{-1}$) & & &  \\
\hline
022233.4+422027	& NGC 891	& 9.1		& $2.40\pm0.01$	& 1(6)	& NGC 891 ULX1	& \cite{earley21}, \cite{hodgeskluck12}, \\
 & & & & & & \cite{wang16}\\
 022727.5+333443	& NGC 925	& 8.7		& $4.5\pm0.3$ (S) 	& 1(1)	& NGC 925 ULX-1	& \cite{salvaggio22}, \cite{pintore18}, \\
 & & & & & & \cite{swartz11}\\
 031819.9-662910	& NGC 1313	& 4.2		& $2.8\pm0.2$ (S)	& 16(29)	& NGC 1313 X-1	& \cite{pinto20}, \cite{bachetti13},\\ 
 & & & & & & \cite{fengandkaaret06}, \cite{colbertandptak02}\\
034615.8+681113	 & IC 342		 & 3.4	 & $1.71\pm0.01$ 	& 1(6)	 & IC 342 X-2	 & \cite{rana15}, \cite{mak11}, \\
 & & & & & & \cite{rlg08}, \cite{ft87} \\
072647.8+854550	 & NGC 2276	 & 39.3	 & $9.9\pm1.5$ (S) 	& 1(1)	 & NGC 2276 3c	 & \cite{mezcua15}, \cite{sutton12}, \\
 & & & & & & \cite{liu11} \\
091948.8-121429	 & PGC 26378	 & 26.5	 & $6.6\pm2.2$	& 1(1)	 & -	 & - \\
 & & & & & & \\
095550.4+694045	 & NGC 3034	 & 3.5	 & $8.50\pm0.02$ 	& 10(10)	 & M82 X-1 	& \cite{brightman20}, \cite{pasham14}, \\
 & & & & & & \cite{kaaret06}, \cite{matsumoto01} \\
112015.7+133514	 & NGC 3628	 & 10.5	 & $3.1\pm0.2$ (S) 	& 1(1)	 & NGC 3628 X-1 & \cite{heil09}, \cite{stobbart06}, \\
 & & & & & & \cite{strickland01} \\
112054.3+531040	 & NGC 3631	 & 20.1	 & $4.4\pm1.3$ (S) 	& 2(4)	 & -	 & \cite{kovlakas20}, \cite{liu11} \\
 & & & & & & \\
131519.5+420301	 & NGC 5055	 & 9.0	 & $6.7\pm1.2$ (S) 	& 2(2)	 & NGC 5055 X-1	 & \cite{mondal20b}, \cite{berghea08}, \\
 & & & & & & \cite{rw00} \\
143242.2-440939	 & NGC 5643	 & 16.1	 & $6.0\pm1.8$ (S) 	& 3(3)	 & NGC 5643 X-1	 & \cite{kosec21}, \cite{krivonos16},  \\
 & & & & & & \cite{pintore16}, \cite{guainazzi04} \\
151558.6+561810	 & NGC 5907	 & 17.1	 & $11.6\pm3.2$ (S)  	& 7(10)	 & NGC 5907 ULX1 & \cite{fuerst23}, \cite{israel17a}, \\
 & & & & & & \cite{walton16}, \cite{sutton13b} \\
165251.5-591503	 & NGC 6221	 & 11.9	 & $3.1\pm0.3$	& 1(1)	 & -	 & - \\
 & & & & & & \\
230457.6+122028	 & NGC 7479	 & 36.8	 & $9.2\pm0.4$	& 1(3)	 & NGC 7479 ULX-1	 & \cite{earnshaw18}, \cite{sutton12} \\
 & & & & & & \cite{walton11} \\
235751.0-323726	 & NGC 7793	 & 3.6	 & $1.5\pm0.1$ (S) 	& 1(9)	 & NGC 7793 P13	 & \cite{fuerst21}, \cite{israel17b}, \\
 & & & & & & \cite{motch14}, \cite{rp99} \\
\hline
\end{tabular} 
\begin{minipage}{0.95\textwidth}
Notes: $^a$ distances are as per adopted by W22.  $^b$ Peak luminosity detected from the ULX, taken from W22.  If this is from a \swift XRT detection we add an (S) to the column; no parentheses indicates an \xmm detection as there were no peak luminosities observed by \chan.  $^c$ Number of times detected in the high flux, high luminosity regime.  Total number of \xmm detections in W22 are given in parentheses.  $^d$ Some examples of nomenclature and prior literature for the ULXs are given, but neither is intended as an exhaustive list.  
\end{minipage}
\end{table*}

\section{Data reduction}
\label{sec:reduction}

\begin{table*}
	\centering
	\caption{Summary of datasets used in this work}
	\label{tab:data}
	\begin{tabular}{lccccc}
		\hline
		ULX	& Mission & ObsID & Date & Exposure $^a$ & Count rate $^b$\\
		(4XMM J)	&	& 	& (YYYY-MM-DD)	& (ks)	& (ct s$^{-1}$) \\
		\hline
		091948.8-121429	& \xmm	& 0694440301	& 2012-06-07	& 3.0/7.4/7.1	& 0.12 $^c$ \\
		 & & & & & \\
		112054.3+531040	& \chan	& 3951	& 2003-07-05	& 89.1	& $48 \times 10^{-3}$ \\
		 & \swift		& $00034428001 \rightarrow 00034428026 ^d$	& 2016-03-15 $\rightarrow$ 2016-10-26	& 42.0	& $(6 - 23) \times 10^{-3}$ \\
		 & \xmm & 0762610401 & 2016-04-18 & 6.2/8.5/8.5 & 0.17\\
		 & \xmm & 0762610701 & 2016-04-20 & 6.2/8.5/8.5 & 0.18 \\
		 & \xmm & 0762610801 & 2016-04-22 & 6.2/8.5/8.5 & 0.17 \\
		 & \xmm & 0762610501 & 2016-05-15 & 18.5/22.4/22.4 & 0.29 \\
		 & \swift 	& $00093203001 \rightarrow 00093203005$	& 2017-09-18 $\rightarrow$ 2017-10-13	& 3.2		& $< 23 \times 10^{-3}$ \\

		 & & & & & \\
		165251.5-591503 $^e$	& \xmm & 0405380201 & 2007-02-16 & 8.7/16.6/16.5 & $< 2.9 \times 10^{-3}$ $^f$ \\
		 & \xmm & 0405380901 & 2007-03-25 & 6.2/7.3/7.5 & 0.13 $^g$ \\
		 & \swift & $00081199002 \rightarrow 00081199004$	& 2016-05-24 $\rightarrow$ 2016-05-26	& 6.9		& $< 8 \times 10^{-3}$ \\ 
		\hline
	\end{tabular}
	\begin{minipage}{0.95\textwidth}
	Notes: $^a$ Detector live time for central chip of \chan and {\it XMM-Newton\/}, or sum of exposures for {\it Swift\/}.  For \xmm this is shown as pn/MOS1/MOS2, and quoted post-filtering for good time intervals (see text).  $^b$ This is shown for \swift as either the range of count rates measured in individual observations (top line), or the upper limit for a sequence with no $3\sigma$ detections (lower lines).  Note two instances of higher $3\sigma$ upper limits -- up to $< 31 \times 10^{-3} \rm ~ct~s^{-1}$ -- were present in observations in the top line.  The \xmm count rates are combined across all three EPIC detectors unless otherwise noted, and corrected to the equivalent on-axis count rate.  $^c$ pn-only count rate.$^d$ There were no observations numbered 00034428006 or 00034428011.  $^e$ The position of \cv was covered by a third observation, 0690580101, but the position of the ULX candidate was at the edge of the field-of-view and no detection of it was present from this dataset in 4XMM-DR10.  We therefore do not analyse this data.  $^f$ $3\sigma$ upper limit for aperture photometry for combined MOS1 and MOS2 data.  $^g$ Combined MOS1 and MOS2 count rate.
	\end{minipage}
\end{table*}

The primary focus of this work were the datasets in which the catalogued ULX detections were made.  These \xmm observations, alongside complementary data from \chan and \swift where analysed, are listed in Table~\ref{tab:data}.  This also provides an indication of the amount of useful exposure per observation, and the count rate detected from each ULX during the observations.

The \xmm data were reduced using the science analysis system (SAS)\footnote{\url{https://www.cosmos.esa.int/web/xmm-newton/sas}} version 20.0.0, with our procedures based largely on the associated analysis threads\footnote{\url{https://www.cosmos.esa.int/web/xmm-newton/sas-threads}}.  All \xmm data in Table~\ref{tab:data} was taken in full-frame mode in all detectors.  We began by reprocessing the data using up-to-date calibration files, and then checked the high energy (10-12\,keV) light curve for each full dataset to determine whether background flare filtering was necessary.  This was not necessary for any of the datasets for \ngcsrc where the full-field count rates were always below the suggested nominal cut-off rates for filtering ($0.35 \ctps$ for the MOS detectors and $0.4 \ctps$ for the pn), with the exception of small number of 100 s pn bins which only marginally exceeded the threshold and so were retained.  In contrast, both datasets for \cv required flare filtering, with the standard thresholds used for observation 0405380201 resulting in the loss of $\sim 10\%$ of the MOS exposure length, and $\sim 50\%$ of that from the pn.  The flaring was much more extreme for 0405380901, where standard filters would have resulted in very little exposure surviving; we therefore adopted much higher threshold for both MOS ($2 \ctps$) and pn ($10 \ctps$), resulting in the loss of $\sim 35\%$ of the original MOS data, and up to $\sim 75\%$ of the pn exposure.  The single exposure covering \pgcsrc was also badly affected by flaring.  After the application of the nominal cut-offs, only $\sim 30\%$ of the MOS exposure was retained, and $< 20\%$ of that from the pn.

A barycentric correction was applied to the event lists.  Light curves and spectra were extracted for \ngcsrc from a 20-arcsecond radius circular aperture centred on the source, with background data extracted from a nearby circular region on the same chip (and at as similar a distance from the readout nodes as possible in pn data) with radius 40 arcseconds.  \cv was 11-12 arcminutes off-axis in each dataset and so larger apertures were used, of 30 arcsecond radius for the source, and 60 arcseconds for the background aperture.  Similar large apertures were also used for \pgcsrc which was even further off-axis.  All light curves were extracted in the 0.2-10\,keV band using events flagged as good (the {\tt \#xmmea$\_$em} and {\tt \#xmmea$\_$ep} selection criteria for the MOS and pn respectively).  We also selected on event pattern, with a threshold of $\leq 4$ for pn data (single and double events) and $\leq 12$ for MOS (which also includes triples and quadruples).  We used {\sc epiclccorr} to subtract background counts and correct the light curves for instrumental effects such as vignetting and deadtime, and where we co-added light curves over the three different detectors we ensured the light curves had the exact same start and stop times (taken from the pn data).  Spectra were extracted using the same patterns and flags as the light curves, excepting that the more stringent {\tt FLAG} $=0$ was used for the pn.  Response matrices (rmf files) and ancillary response files (arf) were produced for each spectrum, with the latter incorporating a correction such the spectra better align with {\it NuSTAR\/} calibration.  Finally, the spectra were binned to a minimum of 25 counts per bin and with an oversampling factor limited to three times the intrinsic energy resolution of the detector across all photon energies.

The \swift data utilised in the analysis of \ngcsrc were all obtained directly from online tools associated with the 2SXPS catalogue\footnote{\url{https://www.swift.ac.uk/2SXPS/}} \citep{evans20}.  The stacked spectrum, composed of all XRT data for the ULX, was obtained directly from the 2SXPS interface, which also provided appropriate background data and response (rmf and arf) files.  The spectral data were binned to 20 counts per bin before fitting.  The \swift data points contributing to the full light curve were extracted on the basis of one point per observation, with the data either displayed as a detection (with corresponding $1\sigma$ error) where the count rate exceeded three times the error, or as a $3\sigma$ upper limit in other cases.

We also include the analysis of an archival \chan dataset for 4XMM J112054.3+531040.  We reduced this data using the \chan interactive analysis of observations software package, {\sc ciao}\footnote{\url{https://cxc.harvard.edu/ciao/index.html}} version 4.14, and version 4.9.8 of the \chan calibration database.  Our reduction again relied heavily on the provided science threads\footnote{\url{https://cxc.harvard.edu/ciao/threads/index.html}}.  The data was reprocessed and data products for the ULX were extracted from an 8-pixel ($\sim 4$ arcsecond) radius aperture centred on the source.  Given the proximity of a second, much fainter source (about 7 arcseconds to the south-west) we chose to use a separate background region rather than an annulus, which we set as a 20-pixel circular aperture to the north-west of the ULX.  We then extracted the source and background spectra, and corresponding response files, using the {\sc specextract} tool, and grouped the output spectrum to a minimum of 20 counts per bin before analysis.  The events file was barycentre-corrected and then background-subtracted light curves were extracted using the same regions and the {\sc dmextract} tool.  Finally, we attempted to produce an improved position for \ngcsrc by correcting the astrometry of the \chan data to corresponding optical counterparts in the {\it Gaia\/} DR2 catalogue (see Section~\ref{ngcsrcoptfup} for more details).

\section{Results}
\label{sec:results}

\subsection{\pgcsrc}

\begin{figure}
\centering
	\includegraphics[width=7.5cm]{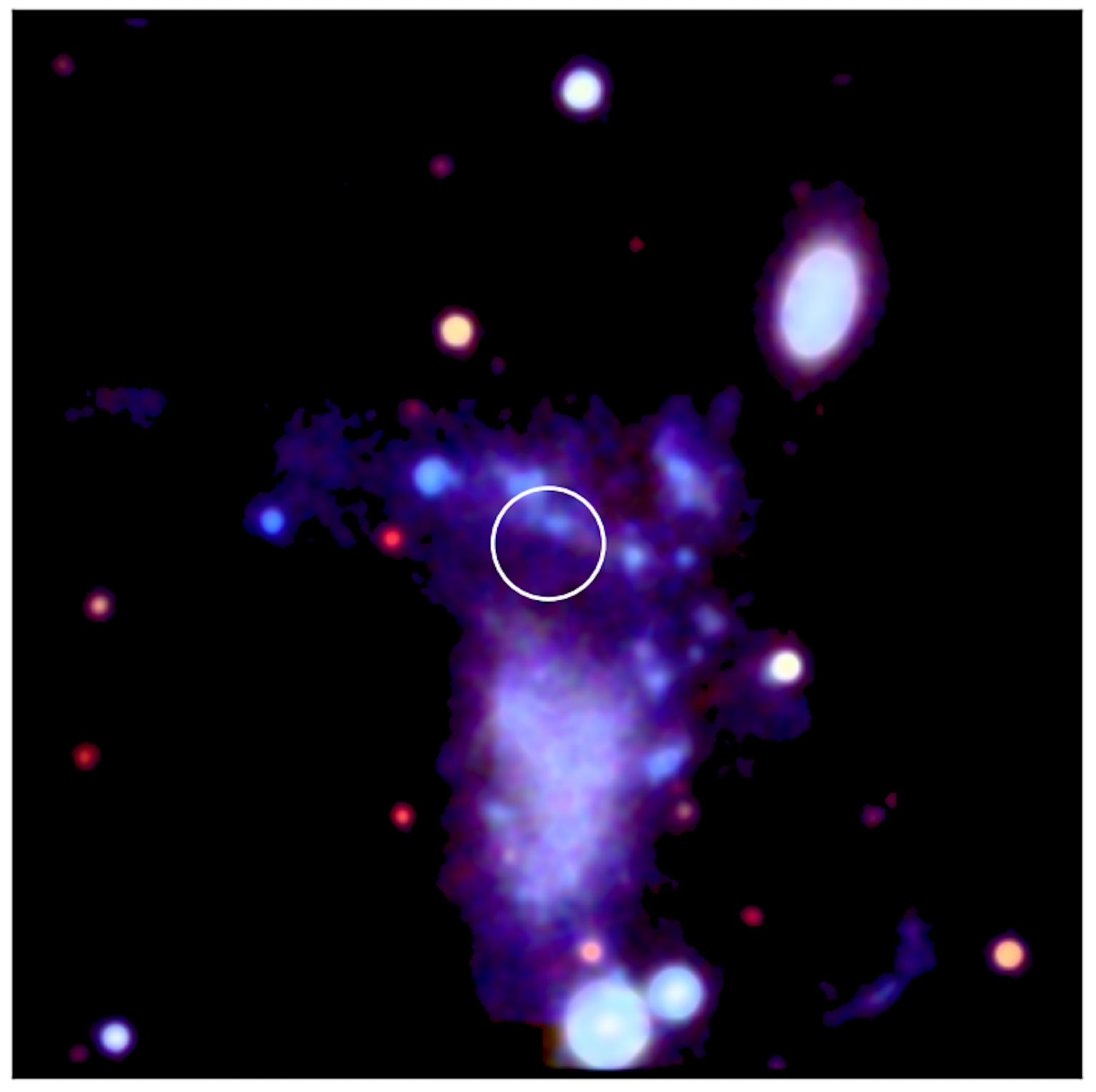}
    \caption{The environment of \pgcsrc shown as a 3-colour PanSTARRs image.  The red, green and blue colours represent light observed in the $i$, $r$ and $g$ filters respectively.  The data in each filter is is convolved with a 2-pixel ($\equiv 0.5$ arcsecond) Gaussian kernel for display purposes.  A circle of radius 6.25 arcseconds marks the $3\sigma$ error in the position of the ULX as determined in 4XMM-DR10, and the field of view is $2\times 2$ arcminutes$^2$.  The PanSTARRs data was obtained from \url{http://ps1images.stsci.edu/cgi-bin/ps1cutouts}.}
    \label{fig:ddo060}
\end{figure}

This candidate ULX is associated with the dwarf galaxy PGC 26378 (also known as DDO 060 or MCG -02-24-011), at a distance of 26.5 Mpc (W22).  It appears co-located with a knot of emission about 30 arcseconds north of the nominal centre of the host galaxy, as shown in Fig.~\ref{fig:ddo060}.  It is only detected by \xmm and inspection of the reduced data of the single observation covering its position reveals it to lie at the very edge of the pn field of view, at $\sim 16$ arcminutes off-axis.  It was not detected in either MOS image given that the 'good' event flags filter used to clean the data reject events so far off-axis.  Hence, only 3 ks of pn data for this source were available for analysis, from which we had $< 100$ source counts\footnote{Given the low detected count rate and severe flaring, taking a higher threshold for the background filtering resulted in background domination at high and low energies, e.g. below 0.5 and above 4 keV for the 10 count s$^{-1}$ GTI filter used for 4XMM J165251.5-591503.  The resulting spectral constraints were no better than those from the method detailed in the text.}.  Strong constraints on the source properties and behaviour are therefore not possible without obtaining further data.

However, we were able to obtain some provisional constraints on the ULX properties by analysing its pn spectrum.  In particular, we extracted the spectrum and then binned it only on the oversampling parameter, such that we did not oversample the energy resolution (precisely as per other \xmm spectra analysed in this work); only we did not specify any minimum number of counts per bin.  This spectrum was then fitted in {\sc xspec}\footnote{Version 12.12.1, available as part of the {\sc heasoft} package from \url{https://heasarc.gsfc.nasa.gov/docs/software/heasoft/}.  Throughout this work we adopt the convention of quoting 90\% errors on spectral fitting results.} using Cash statistics.  We report the results for two simple models -- an absorbed power-law continuum, and an absorbed multi-colour disc blackbody spectrum -- in Table~\ref{tab:pgcspec}.  The fits assumed a Galactic foreground column of $4.85 \times 10^{20} \rm ~cm^{-2}$, interpolated from \cite{dickey90} using the {\sc colden} interface\footnote{\url{https://cxc.harvard.edu/toolkit/colden.jsp}}.  Here and throughout this paper we use the {\sc tbabs} model for absorption \citep{wilms00}.  We find the spectrum to be relatively hard ($\Gamma \sim 1.6$ or $kT_{\rm in} \sim 1.9$ keV), albeit with a very wide range of possible photon indexes or disc temperatures, and with a relatively low intrinsic absorption for a ULX ($\lesssim 6 \times 10^{20} \rm ~cm^{-2}$).  The estimated flux is $\sim 4 \times 10^{-13} \ergcms$, again with a wide uncertainty range, which equates to a 0.3-10\,keV luminosity of $\sim 3 \times 10^{40} \ergs$ at the distance of PGC 26378.

\begin{table}
	\centering
	\caption{Single component spectral model fits for \pgcsrc}
	\label{tab:pgcspec}
	\begin{tabular}{cccc} 
		\hline
		$n_{\rm H}~^a$ & $\Gamma /kT_{\rm in}~^b$ & $f_{\rm X}~^c$ & C-stat/dof$~^d$ \\
		\hline
		\multicolumn{3}{l}{\it power-law continuum} \\
		$0.30^{+0.34}_{-0.25}$ & $1.6^{+0.8}_{-0.6}$ & $4.1^{+2.5}_{-1.6}$ & 29.7/37 \\
		\multicolumn{3}{l}{\it multi-colour disc blackbody} \\
		$< 0.35$ & $1.9^{+4.8}_{-0.8}$  & $3.7^{+1.3}_{-1.4}$& 31.2/37\\		
		\hline
	\end{tabular}
	\begin{minipage}{0.9\columnwidth}
	Notes: $^a$ Absorption in excess of foreground Galactic ($\times 10^{21} \rm~cm^{-2}$).  $^b$ Photon index for power-law, or inner-disc temperature (in keV) for disc blackbody.  $^c$ Observed 0.3-10\,keV flux, corrected for foreground Galactic absorption ($\times 10^{-13} \ergcms$).$^d$ C-statistic value and number of degrees of freedom for fit.  
	\end{minipage}
\end{table}

\subsection{\ngcsrc}

This ULX candidate has the richest dataset of the three we examine in the section, with data from all three missions covered by the W22 catalogue.  We separate out the more detailed analysis this facilitates into three sections below.

\subsubsection{X-ray spectra}

The spectra for \ngcsrc were all extracted and binned as detailed in Section~\ref{sec:reduction}.  After background subtraction they had sufficient counts to permit fitting using the $\chi^2$ statistic.  In this analysis the \xmm and \swift data were fitted in the 0.3-10\,keV band, and the \chan data were fitted in the narrower 0.5-8\,keV regime commensurate with its more limited spectral response.  As an initial step the data were all fitted separately with simple power-law continuum and multi-colour disc blackbody models.  For each spectrum we included a foreground absorption component set to $1.02 \times 10^{20} \pcmsq$, derived from {\sc colden} as above.  We also included a constant component for each \xmm observation to model slight calibration differences between the detectors; in practise differences in this component between detectors never exceeded $15\%$ for any individual observation.  The results of these fits are displayed in Table~\ref{tab:single_spectra}.  It is immediately obvious that this ULX is relatively hard, with a photon index $\Gamma$ that does not deviate substantially from $1.4$, and with moderate absorption beyond our own Galaxy (and so likely to be intrinsic to the ULX or its host galaxy) of $\lesssim 10^{21} \pcmsq$; the equivalent parameters for the disc blackbody model are $kT_{\rm in} \sim 2$ keV and negligible column in excess of Galactic.  The power-law provides a statistically-acceptable fit to all spectra; however this is not the case for all disc-blackbody fits, with \xmm observations 0762610801 and 0762610501 not providing acceptable fits (null hypothesis probability $< 5\%$).  We also show a flux for each observation, which is the observed flux corrected only for foreground Galactic absorption, as calculated using the {\sc cflux} convolution model in {\sc xspec}.  The ULX does appear to vary in flux, albeit only by a factor $\sim 2$ between different observations.

\begin{table}
	\centering
	\caption{Single component spectral model fits}
	\label{tab:single_spectra}
	\begin{tabular}{lcccc} 
		\hline
		ObsID & $n_{\rm H}~^a$ & $\Gamma /kT_{\rm in}~^b$ & $f_{\rm X}~^c$ & $\chi^2$/dof$~^d$ \\
		\hline
		\multicolumn{3}{c}{\it power-law continuum} \\
		3951		& $1.24^{+0.38}_{-0.36}$ & $1.38\pm0.08$ & $5.5\pm0.3$ & 130/144 \\
		\swift stack	& $<1.62$	& $1.21^{+0.26}_{-0.17}$	& $5.4^{+0.8}_{-0.9}$	& 19/16 \\
		0762610401 & $1.30^{+0.84}_{-0.72}$ & $1.57^{+0.19}_{-0.17}$ & $3.6^{+0.4}_{-0.5}$ & 34/27 \\
		0762610701 & $<1.22$ & $1.23^{+0.16}_{-0.15}$ & $4.4^{+0.6}_{-0.5}$ & 16/29 \\
		0762610801 & $<1.13$ & $1.46^{+0.17}_{-0.15}$ & $3.5^{+0.6}_{-0.4}$ & 36/29 \\
		0762610501 & $0.83^{+0.23}_{-0.21}$ & $1.45\pm0.06$ & $6.5^{+0.3}_{-0.5}$ & 128/127 \\
		\multicolumn{3}{c}{\it multi-colour disc blackbody} \\
		3951 	& $< 0.14$ & $1.90^{+0.14}_{-0.12}$ & $4.6\pm0.2$ & 139/144\\		
		\swift stack	& $<0.58$	& $2.09^{+0.62}_{-0.4}$	& $4.4\pm0.7 ^e$	& 17/16 \\	
		0762610401 & $< 0.38$ & $1.52^{+0.22}_{-0.19}$ & $3.0\pm0.4$ & 34/27 \\
		0762610701 & $0^f$ & $2.16^{+0.41}_{-0.3}$ & $3.7\pm0.5$ & 21/30 \\
		0762610801 & $0^f$ & $1.62^{+0.24}_{-0.2}$ & $3.0\pm0.4$ & 50/30 \\
		0762610501 & $< 2.7 \times 10^{-2}$ & $1.87^{+0.11}_{-0.10}$ & $5.8\pm0.3$ & 194/127\\		
		\hline
	\end{tabular}
	\begin{minipage}{0.9\columnwidth}
	Notes: $^a$ Absorption in excess of foreground Galactic ($\times 10^{21} \pcmsq$).  $^b$ Inner disc temperature in keV.  $^c$ Observed 0.3-10\,keV flux, corrected for foreground Galactic absorption ($\times 10^{-13} \ergcms$).  $^d$ $\chi^2$ value and number of degrees of freedom for fit.  $^e$ Absorption fixed at 0 in order to obtain error range.
	\end{minipage}
\end{table}

\begin{figure}
	\includegraphics[width=0.45\textwidth]{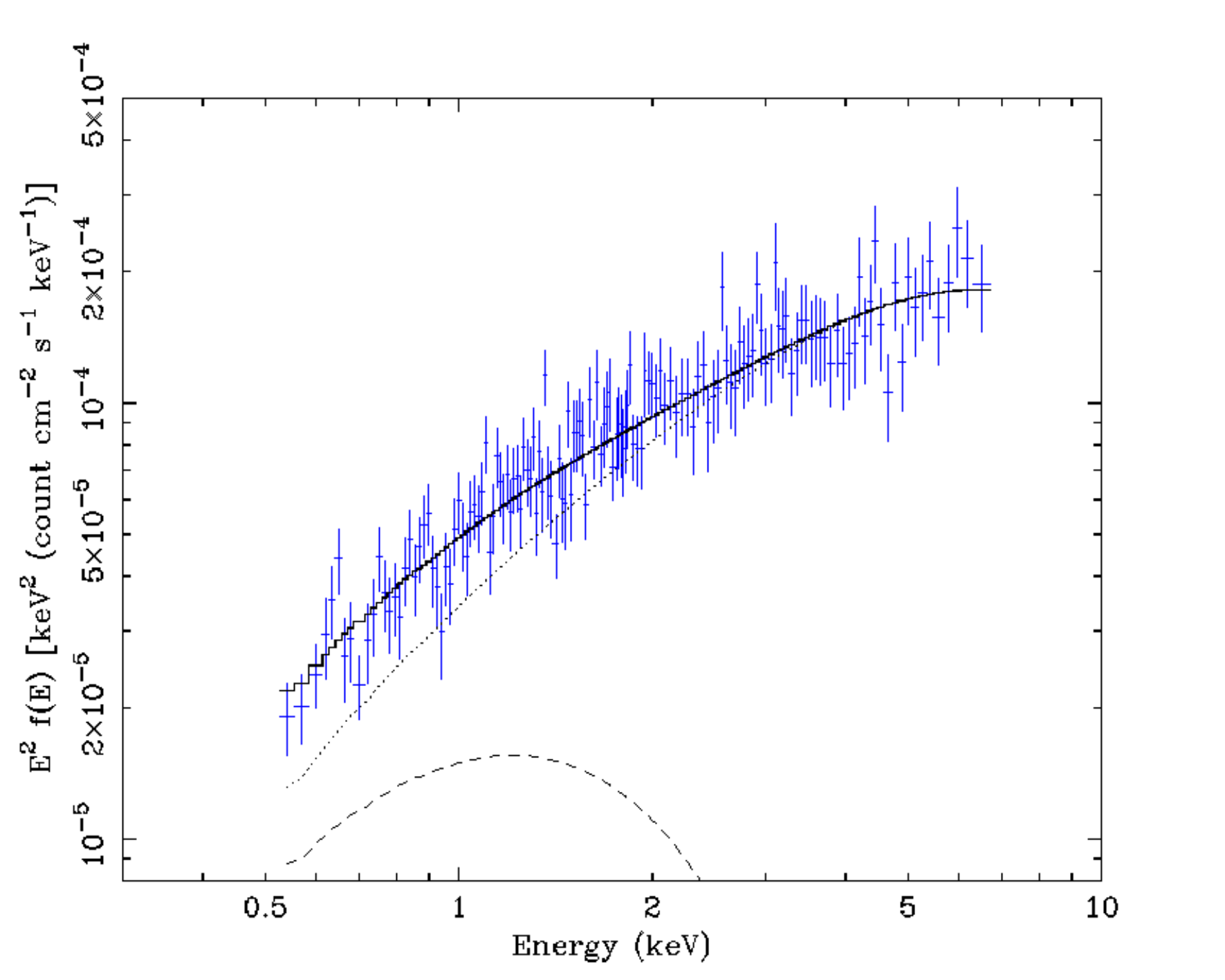}
	\includegraphics[width=0.45\textwidth]{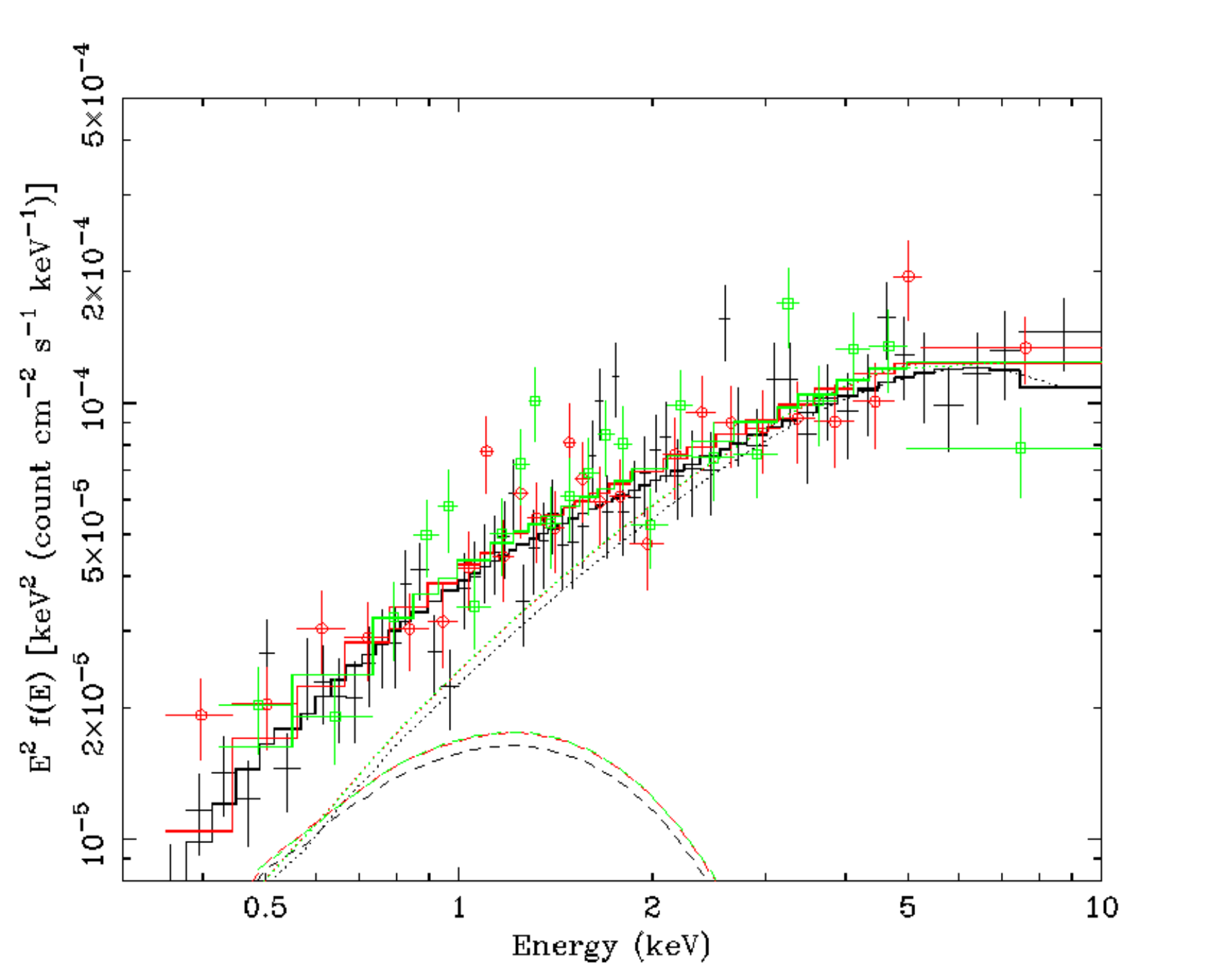}
	\includegraphics[width=0.45\textwidth]{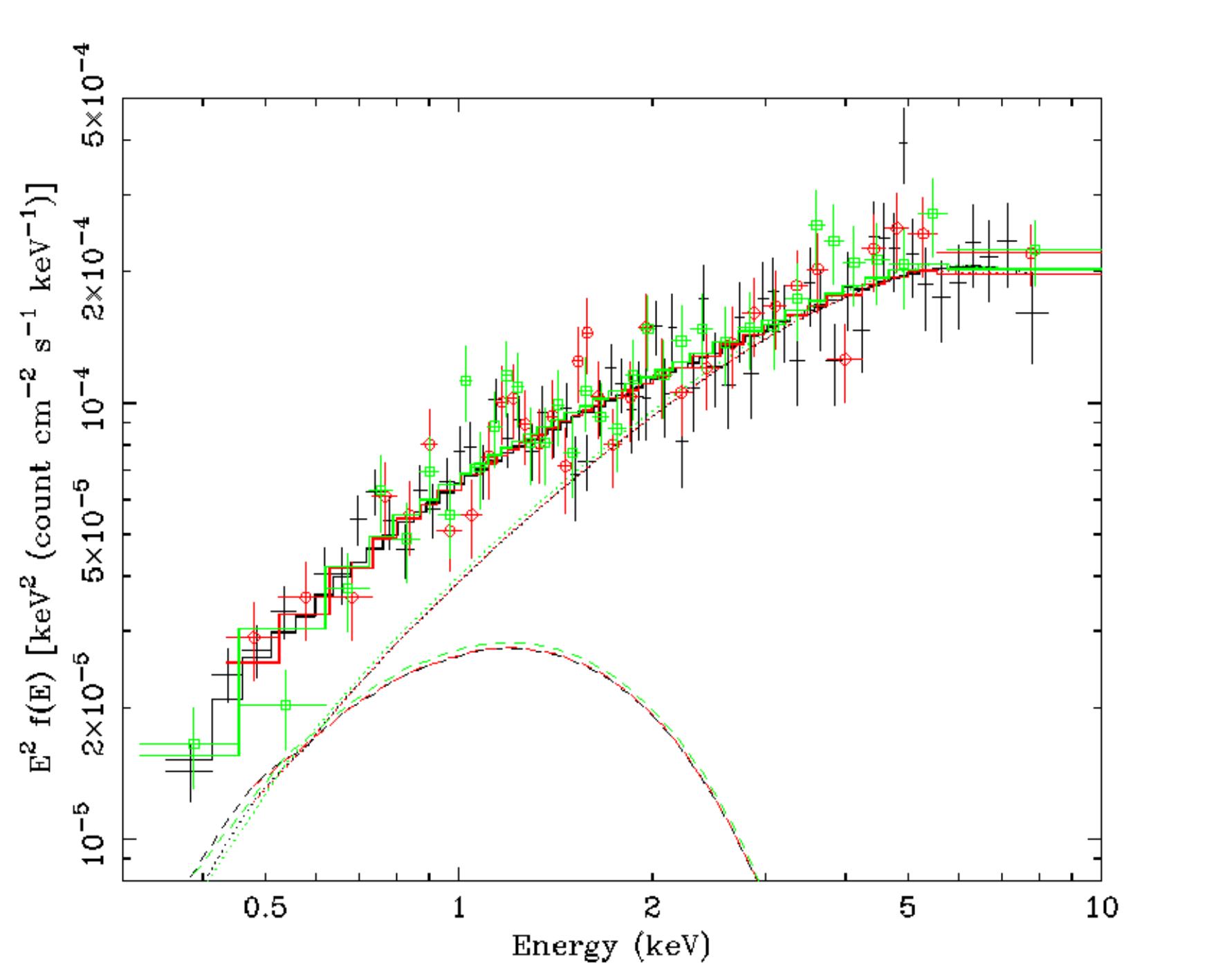}	
    \caption{Unfolded two-component spectral fits for 4XMM J112054.3+531040.  ({\em Top}) \chan spectrum, with data points shown as crosses (in blue).  The best fitting model is the solid line, and the hot and cool disc components are shown as dotted and dashed lines respectively.  ({\em Middle}) Combined spectral data from \xmm observations 07262610401, 0762610701 \& 0762610801.  Model components are indicated as per the top panel (with the slight offsets from the different constants in the modelling).  Data from the pn is shown as simple crosses (in black), MOS1 is highlighted by a circle (red) and MOS 2 by a square (green).  ({\em Bottom}) Data from \xmm observation 0762610501, displayed similarly to the middle panel.}
    \label{fig:N3631_spec}
\end{figure}

Given the similarity of the spectral parameters across all of the X-ray observations, we attempted to fit all the spectra simultaneously with the same power-law model, allowing only the relative constant differences between the \xmm detectors to vary.  This did not produce a good fit to the data when the normalisation of the power-law continua were all constrained to the same value ($\chi^2 = 681.5$ for 391 degrees of freedom).  However, permitting the normalisations to vary (and so the flux of the source to vary between epochs) produced a substantial improvement of $\Delta \chi^2 = 278$ for 5 fewer degrees of freedom, and a statistically-acceptable fit to the data.  Further unfreezing of fit parameters did not result in large statistical improvements -- a small but not very significant improvement of $\Delta \chi^2 = 22$ for 5 fewer degrees of freedom resulted from thawing the photon index, and minimal improvement resulting from thawing the absorption component ($\Delta \chi^2 = 6$ for 5 fewer degrees of freedom).  We therefore conclude that there is no strong evidence for variability in either the absorption component or the spectral shape throughout the observations.  We present the best-fitting power-law and disc-blackbody parameters where absorption and photon index/inner-disc temperature are constrained to the same values across all spectra in Table~\ref{tab:fixedspec}.  Note that the disc-blackbody model is not statistically acceptable, as a result of the two individual spectra noted above; it also does not require any absorption above Galactic foreground in its best fit.

\begin{table}
	\centering
	\caption{Single component fits for \ngcsrc with parameters constrained to the same value across all spectra}
	\label{tab:fixedspec}
	\begin{tabular}{ccc} 
		\hline
		$n_{\rm H}~^a$ & $\Gamma /kT_{\rm in}~^b$ & $\chi^2$dof$~^c$ \\
		\hline
		\multicolumn{3}{l}{\it power-law continuum} \\
		$0.92\pm0.15$ & $1.42\pm0.04$ & 403.6/386 \\
		\multicolumn{3}{l}{\it multi-colour disc blackbody} \\
		$0 ^f$ & $1.86\pm0.07$ & 470.8/387\\		
		\hline
	\end{tabular}
	\begin{minipage}{0.9\columnwidth}
	Notes: $^a$ Absorption in excess of foreground Galactic ($\times 10^{21} \rm~cm^{-2}$).  $^b$ Photon index for power-law, or inner-disc temperature (in keV) for disc blackbody.  $^c$ $\chi^2$ value and number of degrees of freedom for fit.  $^f$ Value fixed to zero as model does not require any additional absorption.
	\end{minipage}
\end{table}

We also consider two-component fits to the spectra, composed of two thermal components, consistent with modelling of good quality 0.3-10\,keV ULX spectra (e.g. \citealt{stobbart06,koliopanos17}).  In order that we can potentially place meaningful constraints we must use only the best quality examples of the spectra of \ngcsrc, which means the \chan spectrum and the 0762610501 \xmm spectrum (both of which have in excess of 100 bins).  In addition, we note that the other three \xmm spectra have similar levels of flux and spectral parameters (Table~\ref{tab:single_spectra}), and were obtained within a 5-day window in 2016 (Table~\ref{tab:data}).  We therefore consider them suitable for combining to a single spectrum per \xmm detector, and to do so used the {\sc SAS} tool {\sc epicspeccombine} to combine the data across the three observations, before binning as previously described.  This had similar spectral quality to the other two datasets and so we used it in the two-component spectral fitting.

Given the similarities of the individual datasets discussed above we also adopted a simultaneous fitting process for the three higher quality spectra.  An initial fit held all values (other than the corrections for relative EPIC detector calibrations) fixed across the three spectra, and a statistically-poor fit resulted $\chi^2 = 652$ for 377 degrees of freedom).  Thawing the normalisations of the hot and cool disk blackbody components both resulted in significant improvements to the fits, with $\Delta\chi^2 = 247$ and 27 (for 2 fewer degrees of freedom) for the hot and cool components respectively when applied consecutively.  However, thawing the disc temperatures and the absorption did not result in any further, additional improvements to the fits ($\Delta\chi^2 \leq 5$ for 2 degrees of freedom), so we keep those values the same across all 3 datasets.  The resulting fit is shown in Table~6, where we also show the intrinsic fluxes of each disc blackbody component (as calculated using {\sc cflux}).  The absorption and disc component temperatures are within the range of values observed from bright ULXs (cf. \citealt{koliopanos17,walton18}), albeit towards the harder end of the range.  The ratio between the flux in either component is rather high, and interestingly may vary between the \chan observation in 2003 and the \xmm observations in 2016, with $f_{\rm X,2}/f_{\rm X,1} = 10\pm4$ in the earlier epoch, and $(6-7)\pm3$ in the latter (note these are $90\%$ and not $1\sigma$ errors).  We illustrate these fits in Fig~\ref{fig:N3631_spec}, where we show the unfolded spectra and both underlying model components.

\begin{table*}
	\centering
	\label{tab:multcompspec}
	\caption{Two component spectral fits for \ngcsrc}
	\begin{tabular}{lcccccc} 
		\hline
		ObsID	& $n_{\rm H}~^a$	& $kT_{\rm in,1}~^b$	& $kT_{\rm in,2}~^b$	& $f_{\rm X,1}~^c$	& $f_{\rm X,2}~^c$	& $\chi^2$/dof $^d$ \\
		\hline
		3951	& &	&	& $0.5\pm0.2$	& $4.8\pm0.2$ 	& \\
		Combined ($401+701+801$)	& $0.29^{+0.29}_{-0.24}$	& $0.48^{+0.17}_{-0.11}$	& $2.69^{+0.54}_{-0.3}$	& $0.5^{+0.2}_{-0.1}$	& $3.2^{+0.2}_{-0.3}$	& 377.8/373 \\
		0762610501	&	&	&	& $0.8^{+0.3}_{-0.2}$	& $5.5\pm0.4$ 	& \\
		\hline
	\end{tabular}
	\begin{minipage}{0.8\textwidth}
	Notes: Where a value is shown solely in the middle row, it is the result of a simultaneous fit to all 3 datasets. $^a$ Absorption in excess of foreground Galactic ($\times 10^{21} \rm~cm^{-2}$).  $^b$ Inner-disc temperature (in keV) for disc blackbody components.  $^c$ Intrinsic flux of component in 0.3-10\,keV band ($\times 10^{-13} \ergcms$); this value is extrapolated beyond the fitted energy range for the \chan data.  $^d$ $\chi^2$ value and number of degrees of freedom for fit. 	
	\end{minipage}
\end{table*}

\subsubsection{X-ray variability}

\begin{figure*}
	\includegraphics[width=\textwidth]{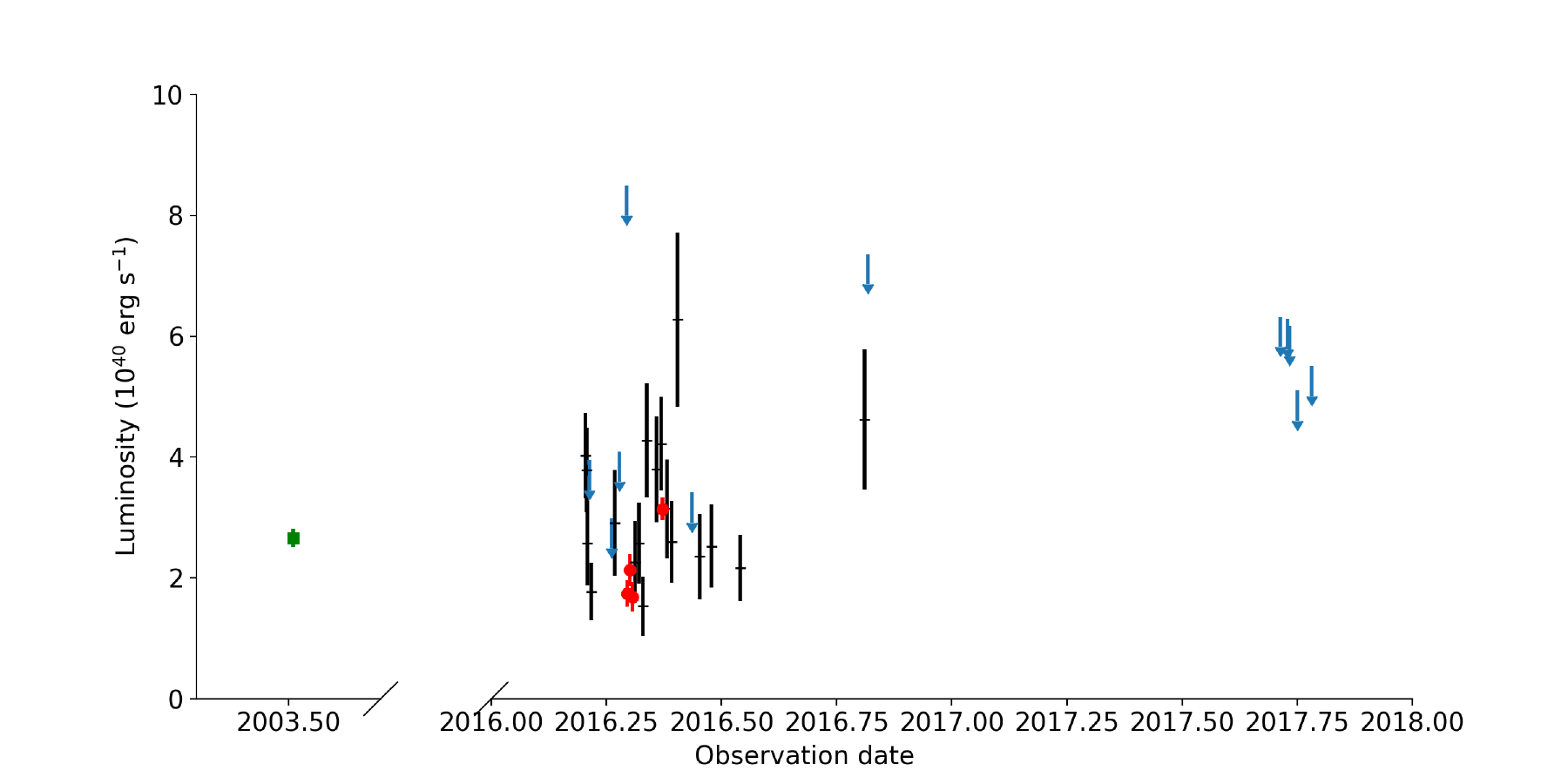}
    \caption{Long-term light curve for \ngcsrc from all missions.  The \chan observation is shown as a (green) square, {\it XMM-Newton} observations are (red) circles, {\it Swift} $3\sigma$ detections are black crosses and $3\sigma$ upper limits from \swift are (blue) downwards arrows.  Note the x-axis is broken for display purposes, given the long (13-year) gap between the \chan and subsequent observations.}
    \label{fig:N3631_ltlc}
\end{figure*}

We have composed a long-term light curve for \ngcsrc from the archival {\it XMM-Newton\/}, \chan and \swift data, which we display in Fig~\ref{fig:N3631_ltlc}.  The \xmm and \chan fluxes were taken directly from the individual power-law spectral fits and converted to a luminosity for a distance of 20.1 Mpc; the \swift fluxes were calculated based on a conversion factor of $1 \ctps \equiv 5.65 \times 10^{-11} \ergcms$, derived from the best fit spectral model in the {\sc pimms}\footnote{\url{https://cxc.harvard.edu/toolkit/pimms.jsp}} calculator.  The light curve shows that \ngcsrc is moderately variable on timescales of days (up to factors of $\sim 5$ in \swift data), but has remained luminous in all observations to date with $L_{\rm X}$ (0.3-10\,keV) persistently $> 10^{40} \ergs$.  However, coverage is limited, with most observations in a short window in 2016 as the \xmm and \swift observations were a follow-up programme for a supernova, AT2016bau (see \citealt{arbour16, granata16}).  The supernova lies $\sim 45$ arcseconds from the ULX, in the direction of the centre of the host galaxy (cf. Figure~\ref{fig:HST_figure}), and remained undetected in the \xmm and \swift data.  We note the only period of observations without a detection were the five follow-up observations with \swift in late 2017.  These observations were short and so do not provide stringent individual limits, however combining their data provides a $3\sigma$ upper limit on the luminosity of $< 2.05 \times 10^{40} \ergs$ for that period, similar to the lowest detected luminosities from a year earlier.

\begin{figure*}
	\includegraphics[width=\textwidth]{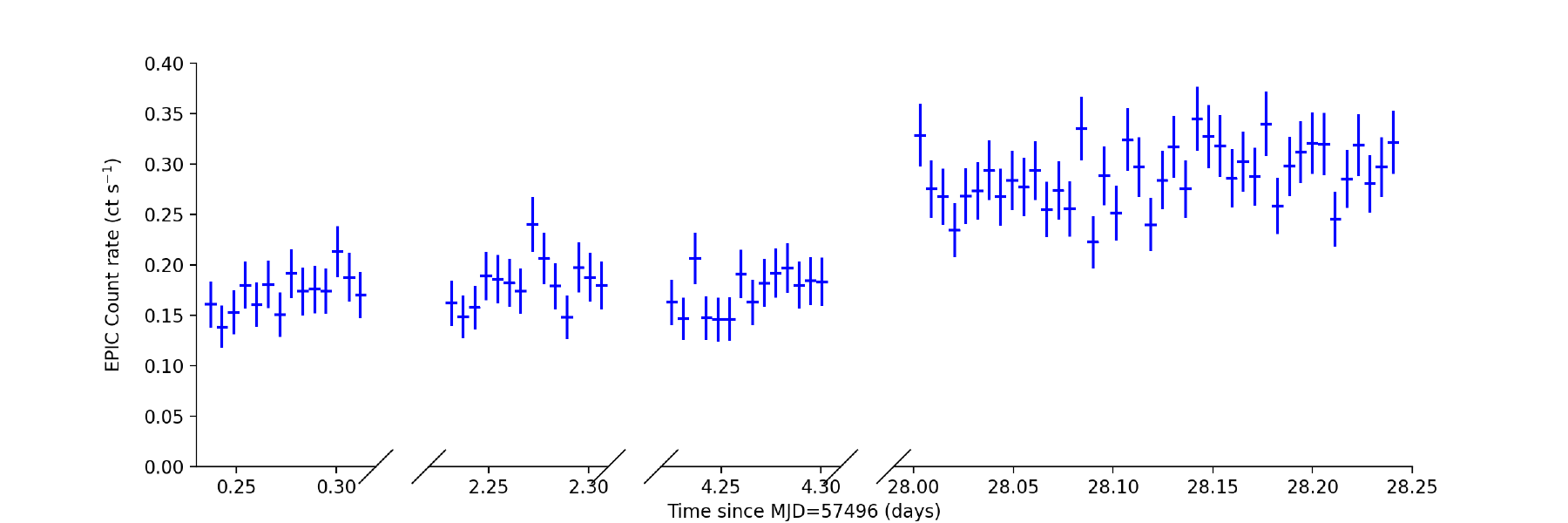}
    \caption{Combined EPIC light curves for \ngcsrc from each individual exposure, displayed with 500 s binning.  Note the x-axis is broken for display purposes.}
    \label{fig:N3631_stlc}
\end{figure*}

\begin{figure}
	\includegraphics[width=0.45\textwidth]{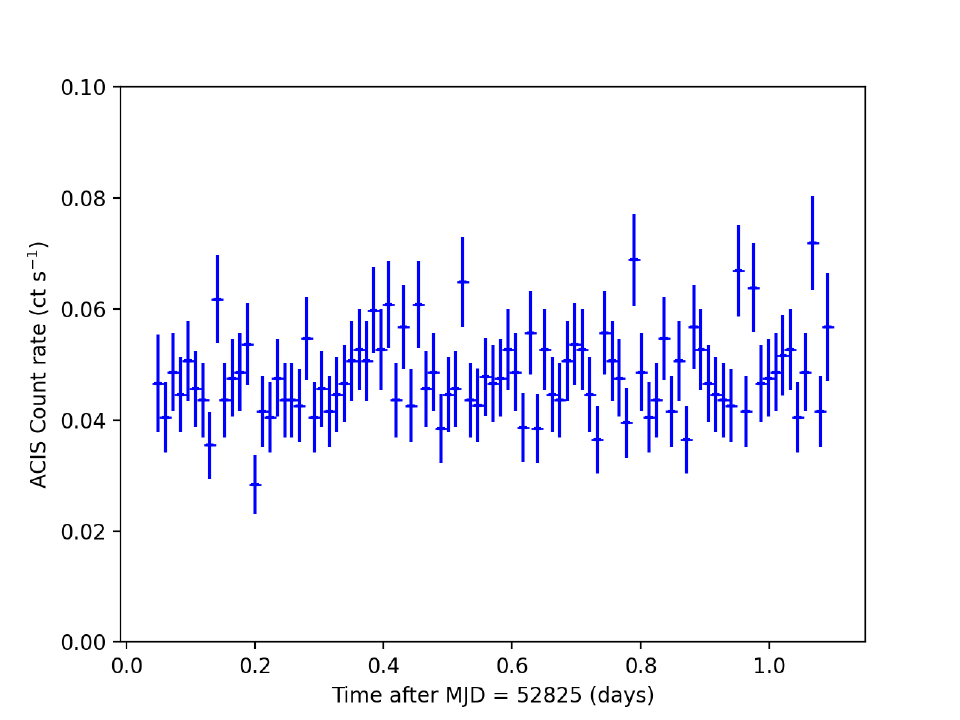}
    \caption{\chan ACIS-S light curve for \ngcsrc shown with 1000 s binning.}
    \label{fig:N3631_stlc_chan}
\end{figure}

We also investigated the intra-observational variability in the \xmm and \chan data.  In Fig~\ref{fig:N3631_stlc} we show summed EPIC light curves for all four \xmm observations.  In each case we limit the light curves to the pn start and stop times to ensure simultaneity across the three EPIC detectors, and bin the data to 500 s intervals.  Visual inspection of Fig~\ref{fig:N3631_stlc} hints that the data might vary in excess of that expected from white noise processes, and this is confirmed by calculating the excess variance ($\sigma_{\rm XS}$; see \citealt{vaughan03}) for each dataset, which in each case is in the range $4-9\%$.  We performed the same analysis on the \chan dataset, binned to 1000 s given its lower count rate, and found a similar $\sigma_{\rm XS} = 7\%$; we show this light curve in Fig.~\ref{fig:N3631_stlc_chan}.  This demonstrates that \ngcsrc is consistently varying on timescales as short as $< 10$ minutes.

In order to investigate variability at the shortest accessible timescales we used the the {\sc stingray} packages \citep{huppenkothen19,bachetti22}, as scripted up in the {\sc hendrics} tools\footnote{Version 7.0, see \url{https://hendrics.stingray.science/en/stable/index.html}.}.  We worked on the \xmm EPIC-pn data as its 73.4 ms frame time means it is the only data for \ngcsrc with the time resolution adequate to access the $\sim 1$ Hz pulsations seen in PULXs.  We extracted pn event files from the source apertures for each \xmm observation, and performed barycentric corrections on each event list.  We first used these events files to create power spectra for each observation.  Each power spectrum was subject to geometric rebinning over a variety of factors in the range 0.01 -- 0.3 before inspection; however, there was no power above the white noise level in any of the power spectra, which were sensitive to power in the $\sim 10^{-3} - 6$ Hz range.  (We note this is consistent with the excess variance results above, which are based on variability over longer timescales than probed by the power spectra.)  We then performed accelerated pulsation searches, using the {\sc hendrics} implementation of the \cite{ransom02} algorithm.  Although each search found candidate pulsations, these were statistically weak (highest powers in the range 30 -- 40), and were generally not recovered by $Z^2_N$ searches focussing on a narrow range around the candidate frequency, so we do not regard any as real pulsation candidates.  This $Z^2_N$ test was used to place limits on the amplitude of possible pulsations in the data; for the shorter ($\sim 6$ ks in the pn) observations the $3\sigma$ upper limit on the possible pulsation amplitude at frequencies $\gtrsim 1$ Hz was $\sim 45\%$, and for the longer ($18$ ks, 0762610501 dataset) a more stringent limit on the pulsation amplitude of $\sim 20\%$ at similar frequencies was calculated.

\subsubsection{Optical follow-up}
\label{ngcsrcoptfup}

An accurate position for \ngcsrc was obtained from the \chan data by considering matches between X-ray source detections on the S3 chip and optical sources from {\it Gaia\/} DR2, as per the relevant science thread\footnote{\url{https://cxc.cfa.harvard.edu/ciao/threads/reproject_aspect/}}.  The best solution was obtained from the raw data; any attempts to correct the astrometry algorithmically induced false matches within the body of NGC 3631, which resulted in no good matches in the field and an overall matching error $> 0.6$ arcseconds.  If we used the raw data we instead found three excellent matches and a statistical error of $0.14$ arcseconds on the field astrometry.  Using this astrometric solution we find a J2000 position for the eULX of $11~20~54.306, +53~10~40.68$ $\pm 0.02$~arcseconds (X-ray centroid error) $\pm0.14$~arcseconds (astrometric uncertainty).

\begin{figure}
        \centering
	\includegraphics[width=0.45\textwidth]{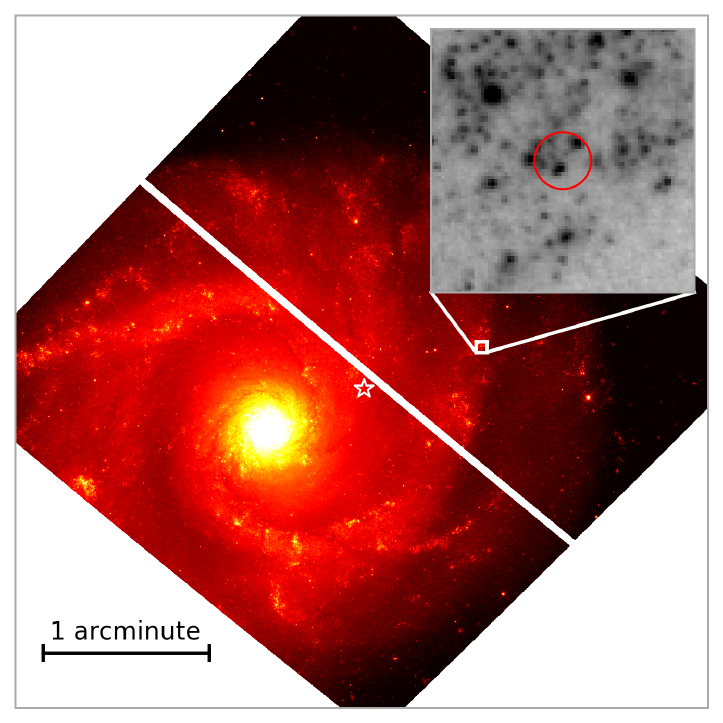}
    \caption{\hst ACS WFC image of NGC 3631 in the F814W filter (main image, using a heat colour map).  The image is aligned such that North is up, and East to the left.  The location of \ngcsrc is within the small white box, and the region within the box is shown in detail in the insetted zoom, which has size $4 \times 4$ arcseconds$^2$ and is displayed as a greyscale image.  The position of the eULX is highlighted by the red circle, which is equivalent to the $3\sigma$ error region for the eULX position.  The position of SN2016bau is highlighted by the white star.}
    \label{fig:HST_figure}
\end{figure}

We then investigated the environment and possible counterparts of \ngcsrc using \hst data.  Specifically, the eULX position was covered by three \hst images.  A pair of WFC3 UVIS images, in the F555W and F814W filters, each of $\sim 750$~s exposure (datasets ID9610010 \& ID9610020), were taken in October 2016 and aimed at SN AT2016bau, as per most of the X-ray data; and a deeper $\sim 2300$~s ACS WFC image was taken in April 2019, also in the F814W filter (dataset JDXK07010).  We display the latter, with the ULX position highlighted in a zoom of its projected immediate vicinity, in Fig~\ref{fig:HST_figure}.  \ngcsrc clearly lies in a spiral arm to the west of the nucleus of NGC 3631, and is associated with some structure within the arm, consistent with it being a {\it bona fide\/} eULX rather than a background AGN.

Fig~\ref{fig:HST_figure} shows two relatively bright sources within the $3\sigma$ uncertainty region, and a third just outside.  The central source appears potentially extended towards the south-east, so may be an amalgamation of two (or more) objects.  The Hubble source catalog\footnote{\url{https://catalogs.mast.stsci.edu/hsc/}} provides magnitudes in both filters from the 2016 WFC3 UVIS observation; both objects within the error region have near-identical magnitudes of $m_{\rm F555W} = 23.86/23.87$ and $m_{\rm F814W} = 23.68/23.66$, for the central and north-western objects respectively, and hence both have a colour $m_{\rm F555W} - m_{\rm F814W} \approx 0.2$.  The object to the east of the error region is very slightly brighter, with $m_{\rm F555W} = 23.36$, $m_{\rm F814W} = 23.51$ and so a colour of $m_{\rm F555W} - m_{\rm F814W} = -0.15$.  After correcting for minimal foreground extinction from within our own Galaxy, absolute magnitudes of $M_{\rm F555W} \approx -7.7$ and $M_{\rm F814W} \approx -7.5$ can be calculated for the two most likely counterparts.  This combination of absolute magnitude and slightly red colour is not a good match for any type of supergiant star, but may perhaps instead be indicative of young stellar clusters in NGC 3631, with the possibility of some reddening due to local dust extinction in the spiral arm.  It is also likely that the observed magnitudes of these counterparts rule out a background AGN, given these typically have $f_{\rm X}/f_{\rm opt} \sim 0.1 - 10$ (e.g., \citealt{aird10}), whereas $f_{\rm X}/f_{\rm opt} > 220$ for this eULX.

\subsection{4XMM J165251.5-591503}

The position of \cv has been covered by 3 \xmm observations, but only one (ObsID 0405380901) provided a reasonably high source count rate and this observation was heavily affected by flaring (cf. Table~\ref{tab:data} and Section~\ref{sec:reduction}).  Worse still, the source lay at the very edge of the pn detector where a sizeable fraction of the source extraction aperture lay outside the field-of-view; we therefore concentrated our analysis on the MOS detections of this source, where the flare contamination was also less pronounced than for the pn.  Spectra were extracted and fitted with simple single component models, similarly to the previous sources.  We report the results of this analysis in Table~\ref{tab:cvspec}.  In this case the foreground column was higher, at $1.53 \times 10^{21} \pcmsq$.  We find the source to display a relatively soft spectrum, with $\Gamma \sim 2.5$ for a power-law continuum, or $kT_{\rm in} \sim 0.65$~keV for a multi-colour disc blackbody model.  The latter shows no evidence for absorption above the Galactic line-of-sight.  Both models provide formally acceptable fits to the data, although the $\chi^2$ is better for the power-law model.  The source flux (corrected for Galactic absorption) is the highest for any of our three new eULX candidates at $\sim 10^{-12} \ergcms$.  However, this is not a persistent flux; we note that the upper limit on the count rate from a second observation (taken roughly 5 weeks prior to the dataset fitted above) was a factor $\sim 45$ lower, so this object is strongly X-ray variable on a timescale of weeks.  

This variability is corroborated by \swift data; the position of \cv is covered by three \swift observations (as listed in Table~\ref{tab:data}, the target for which was NGC 6221), taken over 3 days in 2016 and totalling 6.9 ks of exposure.  No source is detected at the ULX position in 2SXPS analysis of this data.  We have therefore performed aperture photometry at the position of the source and find a combined $3\sigma$ upper limit on the flux as $\lesssim 2 \times 10^{-13} \ergcms$ (converting the upper limits on the count rate to flux in \textsc{pimms} using the power-law spectrum in Table~\ref{tab:cvspec}).  Although not as constraining as the second \xmm observation, this again highlights that this object is variable and, nine years after the original detection, it had a factor $\geq 5$ lower flux compared to the earlier epoch.

\begin{table}
	\centering
	\caption{Single component fits for \cv}
	\label{tab:cvspec}
	\begin{tabular}{cccc} 
		\hline
		$n_{\rm H}~^a$ & $\Gamma /kT_{\rm in}~^b$ & $f_{\rm X}~^c$	& $\chi^2$dof$~^d$ \\
		\hline
		\multicolumn{3}{l}{\it power-law continuum} \\
		$2.0_{-1.2}^{+1.5}$ & $2.5_{-0.4}^{+0.5}$ & $1.1_{-0.2}^{+0.3}$	& 54.4/50 \\
		\multicolumn{3}{l}{\it multi-colour disc blackbody} \\
		$0 ^f$ & $0.65_{-0.1}^{+0.13}$ & $0.9_{-0.2}^{+0.1}$	& 66.2/51\\		
		\hline
	\end{tabular}
	\begin{minipage}{0.9\columnwidth}
	Notes: $^a$ Absorption in excess of foreground Galactic ($\times 10^{21} \rm~cm^{-2}$).  $^b$ Photon index for power-law, or inner-disc temperature (in keV) for disc blackbody.  $^c$ Observed 0.3-10\,keV flux, corrected for foreground Galactic absorption ($\times 10^{-12} \ergcms$).  $^d$ $\chi^2$ value and number of degrees of freedom for fit.  $^f$ Value fixed to zero as model does not require any additional absorption.
	\end{minipage}
\end{table}

The high foreground column is indicative that the line-of-sight to this eULX candidate lies close to the Galactic plane.  In fact, its Galactic coordinates are $l_{\rm II} = 329.72, b_{\rm II} = -9.60$, which lies both close to the Galactic plane, and near to the direction of the Galactic centre.  This, combined with the soft spectrum, lead to a suspicion that this could be a foreground Galactic source.  We therefore conducted a more extensive check for multi-wavelength counterparts and found that \cv has a bright IR counterpart, detected within $\sim 1$ arcsecond of the eULX candidate position, with $m_{\rm K} = 13.01\pm 0.03$ in the 2MASS all-sky point source catalogue \citep{cutri03}.  Given this strongly suggests a Galactic counterpart we also checked {\it Gaia\/}-DR3 \citep{gaia22}, and found the counterpart with $m_{g,\rm mean} = 15.6$ and a parallax of $0.41\pm 0.03$  milli-arcseconds.  The presence of a parallax means that this counterpart is clearly Galactic, and the measured value places it at a distance of $2.4 \pm 0.2$ kpc.  The apparent magnitude of the counterpart then converts to a mean absolute magnitude of $M_g \approx 3.2$, correcting for foreground extinction of $A_g = 0.5$ (\citealt{schlafly11}, via the NASA/IPAC extragalactic database\footnote{\url{https://ned.ipac.caltech.edu}}), and the 0.3-10\,keV X-ray luminosity is $L_{\rm X} \approx 7 \times 10^{32} \ergs$.

\begin{figure*}
	\includegraphics[width=0.8\textwidth]{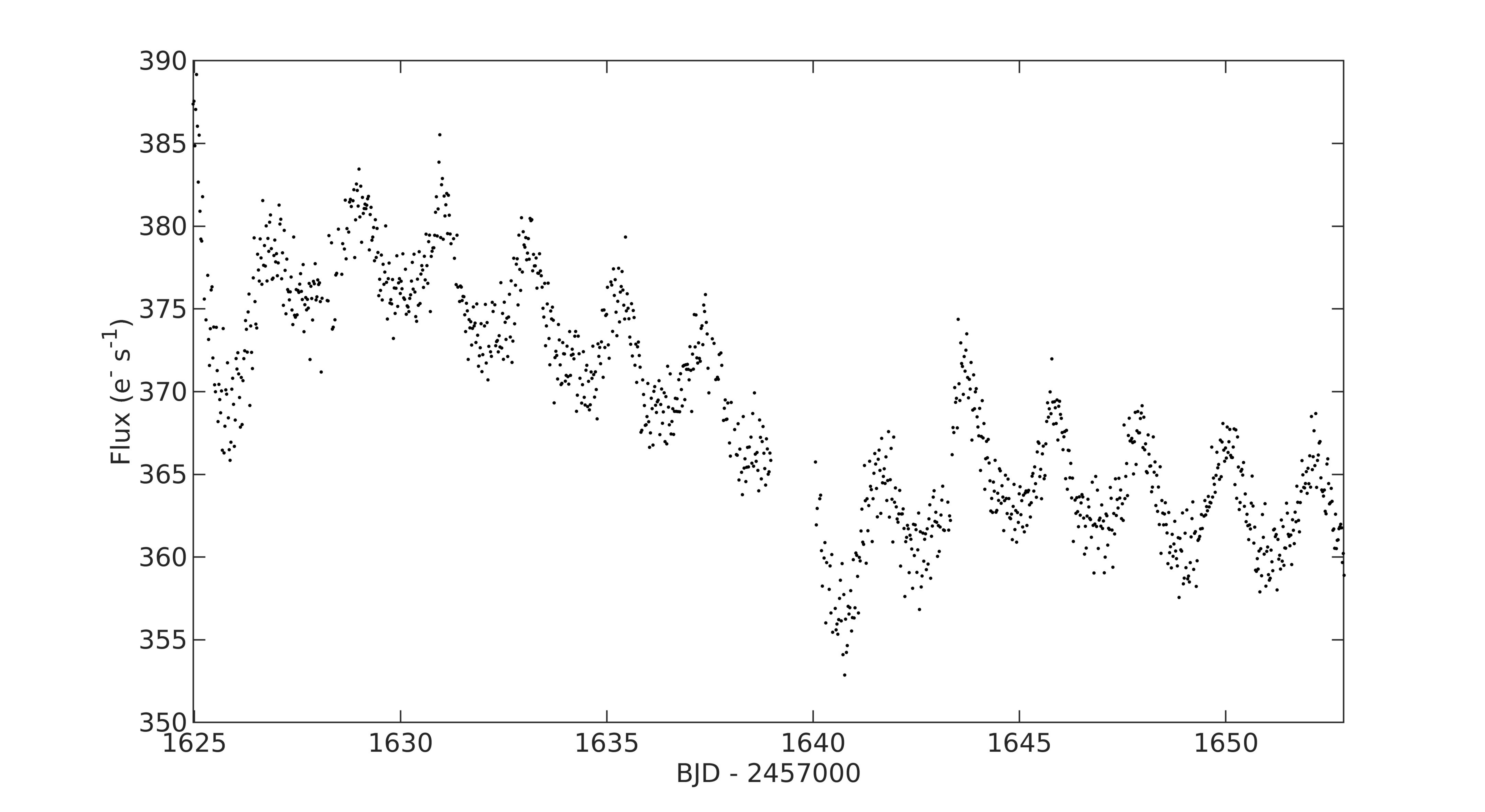}
    \caption{Light curve of the candidate counterpart to \cv obtained with 30 minute cadence during Sector 12 of the {\it TESS\/} mission.}
    \label{fig:TESS_lc}
\end{figure*}

To investigate its nature further we extracted a {\it TESS\/} light curve at its position.   {\it TESS\/} observed the target during Sector 12 (2019-May-21 to 2019-Jun-19) at 30 minute cadence. We extracted its light curve by defining target and background apertures around the target and show the resulting light curve in Fig~\ref{fig:TESS_lc}. The counterpart displays a clear $\sim 2$ day modulation, on top of a possible longer term variation.   Given the {\it Gaia\/} distance and colour, the position in the {\it Gaia\/} calibrated colour-magnitude diagram places this system as a K-type star, with its absolute magnitude suggesting an evolved type.  It may therefore be possible that the 2-day modulation is related to the orbital period of a binary, with the X-ray emission potentially originating in colliding winds.  Alternatively the optical variability could be related to a super-orbital modulation caused by a precessing tilted disk as observed in several cataclysmic variables (see e.g. \citealt{ilkiewicz21}); however it is unclear what the origin of the X-ray emission might be in this scenario.  A third possible scenario is unrelated to the possible binary nature of the stellar system; this may simply be a stellar flare.  At close to $10^{33} \ergs$ this would be at the upper end of known flare luminosities (cf. Table 4 of \citealt{gudel04}), and the ratio of quiescent to flare luminosity would also be rather extreme for a late stellar type \citep{pye15}, but this is a plausible nature given its soft X-ray spectrum.  It is even possible that this optical counterpart is entirely unrelated to the X-ray emission, which still might be originating in a {\it bona fide\/} ULX in NGC 6221.  The nature of this source therefore remains to be confirmed and will require follow-up observations to resolve.

\section{Discussion}
\label{sec:discussion}

\subsection{Have we found any good PULX candidates?}

In this paper we have identified and studied three new extreme ULX candidates, selected on the basis of at least one high flux ($>5 \times 10^{-13} \ergcms$) and high luminosity ($L_{\rm X} > 10^{40} \ergs$) \xmm detection in the ULX catalogue of W22, with a view to identifying good PULX candidates.  So, what have we learnt?

The first object, 4XMM J091948.8-121429, has insufficient data to determine whether it is a good candidate to be a PULX.  Its host is relatively distant at 26.5 Mpc, and its only detection has it 16 arcminutes off-axis in 3 ks of cleaned pn data, so $< 100$ counts were obtained from the source.  This was however sufficient to determine the object appears likely to be within its host given its association with structure in PGC 26378, and that its spectrum is likely to be moderately hard (although this is poorly constrained).  It is pertinent to note that the host galaxy is a dwarf irregular system; several such systems are known to host eULXs, including some of the best studied local objects (e.g. NGC 5408, Holmberg II).  Indeed some of these objects are particularly interesting as local analogues for processes at high redshift, e.g. I Zw 18 \citep{kaaret13} and Haro II \citep{prestwich15}.  So, \pgcsrc is at least a good candidate for a {\it bona fide\/} eULX, but further observations are required to reveal the details of this interesting object.  

4XMM J112054.3+531040, in contrast, has a reasonable set of data already available, largely thanks to follow-up of a supernova that detonated in the host galaxy, NGC 3631, in 2016.  This includes over 40 ks of \swift snapshots, four short \xmm observations, and a much earlier (2003) and deeper \chan observation.  These reveal a source that varies in flux by a factor $\sim 2-3$ on timescales from days to years, and that has demonstrable variability on timescales as short as tens of minutes.  Its spectrum is generally hard for a ULX, represented by a power-law with $\Gamma \sim 1.4$, however it can also be satisfactorily described by a two component model with both components being thermal in nature, with the form and its parameterisation within the known range of spectra for ULXs.  Combining \chan astrometry with \hst imaging reveals possible counterparts to the ULX within a spiral arm of the host, that have colour and magnitude similar to young stellar clusters, but far too high an X-ray/optical flux ratio for a background AGN.  It therefore is a good eULX detection, and its hardness marks it out as good PULX candidate given that PULXs tend to have harder-than-average X-ray spectra amongst ULXs, generally due to the dominance of a hard and variable spectral component in the 0.3-10\,keV range \citep{walton18,gurpide21}.  We do not detect pulsations in any individual \xmm observation, with only one relatively stringent limit of $\lesssim 20\%$ at frequencies $> 1$ Hz.  Even then, we note that pulsations are a transient phenomenon in PULXs and are not seen in all epochs; and indeed the pulsed fraction itself varies in the 0.3-10\,keV band, with it dropping below the best limit seen for \ngcsrc in several observations of NGC 7793 P13 \citep{fuerst21}, or never exceeding 10\% in the faintest of the detected pulsations from NGC 1313 X-2 \citep{sathyaprakash19}.  It is therefore still eminently possible that further observations of \ngcsrc could reveal pulsations, and we regard it as a good candidate PULX.

The final object, 4XMM J165251.5-591503, has been shown to likely be a foreground object within our own Galaxy, although its precise nature is not clear. The fact it was not removed from W22 during the composition of that catalogue shows a potential weakness of its selection/rejection criteria -- a lack of detailed consideration of Galactic contaminants.  Although stars are rejected if spatially coincident with candidate ULXs, this is limited to those in the Tycho 2 catalogue \citep{hog00} and those that are picked up via the Simbad interface\footnote{\url{http://simbad.cds.unistra.fr/simbad/}}.  This example shows that more filtering is required to pick up Galactic objects, for example looking for counterparts with measured parallax at low Galactic latitudes.  The 4XMM-DR9 ULX catalogue of \cite{bernadich22} performs better than W22 in this regard by searching catalogues including {\it Gaia\/} DR2 for optical point source counterparts to its ULX candidates, and then using the X-ray/optical flux ratios to exclude stars and other objects.  The even more recent catalogue of \cite{tranin23} uses an automated probabilistic classification of X-ray sources to produce a catalogue with a claimed contamination of only 2\%.   Such stratagems will help to improve future generations of ULX catalogues.

\subsection{Are there alternative targets for study from \chan or \swift?}

In total, our selection strategy revealed three previously unstudied eULXs from a sample of 15 in W22 that had relatively high flux and high luminosity detections in \xmm data.  The focus on \xmm was specifically so that we could select pre-existing data where pulsation searches were possible at frequencies similar to those of the pulsations of known PULXs, which is currently only possible below 10 keV using pn data.  It is interesting to ask whether the other mission catalogues analysed in W22 revealed broadly similar eULX candidates; although they might not have \xmm data suitable for sensitive pulsation analysis currently available, they would potentially be excellent targets for future observations.  We therefore repeated the selection of eULX candidates using the same two simple flux and luminosity criteria, but using the ULX catalogues in W22 derived from \swift (2SXPS) and \chan (CSC2.0).

\begin{figure*}
	\includegraphics[width=0.48\textwidth]{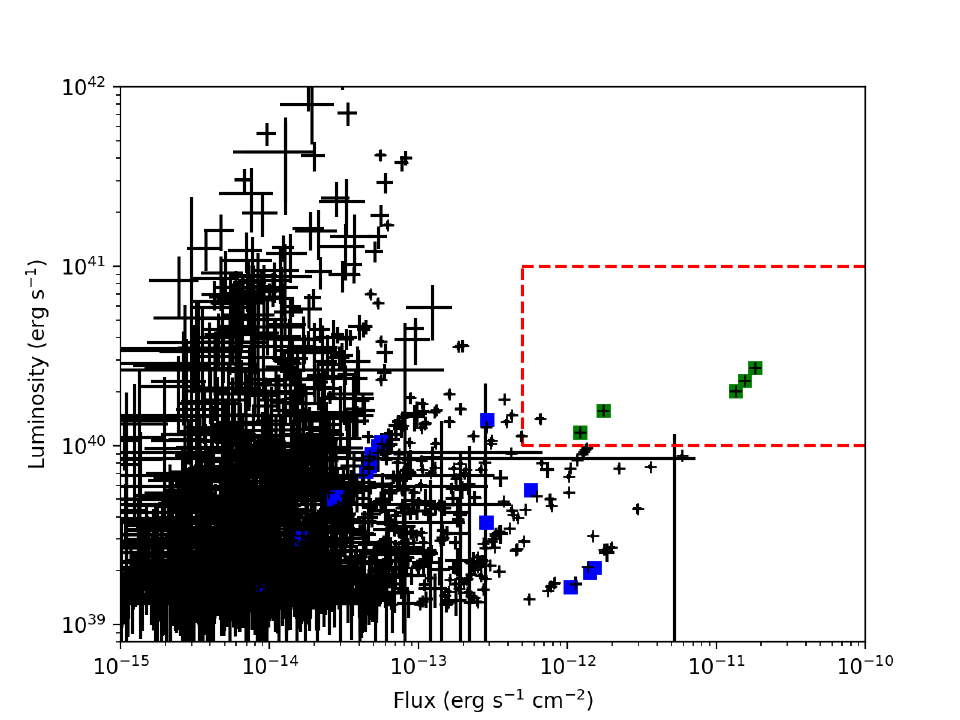}
	\hspace{0.5cm}
	\includegraphics[width=0.48\textwidth]{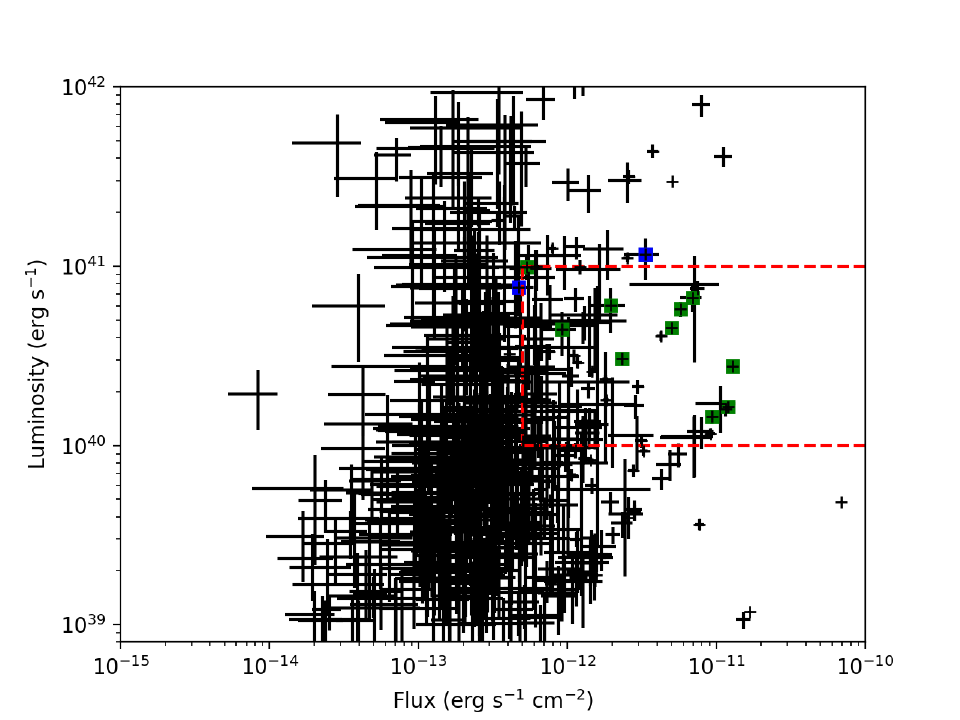}
    \caption{Re-creation of Fig~\ref{fig:source_select}, but instead for the CSC2.0 ({\it left\/}) and 2SXPS ({\it right}) ULX candidates from W22.  Any of the 15 sources detected by \xmm within the region of interest (delineated by the red, dotted line) also detected in the same region of parameter space by \chan or \swift are highlighted by a green square.  Any of these objects detected but not in the region of interest are highlighted instead by blue squares.  The CSC2.0 fluxes and luminosities are all increased by a factor 1.3 to account for its smaller bandpass; the 2SXPS data panel uses the peak values for each individual source (i.e. the highest flux/luminosity combination) from any individual \swift observation.}
    \label{fig:altsrc_select}
\end{figure*}

The results are shown in Fig~\ref{fig:altsrc_select}.  To create this figure we needed to correct the CSC2.0 data for its narrower (0.5-7\,keV) bandpass.  For a variety of spectral shapes representative of most ULXs (power-law continua in the range $\Gamma \sim 1.5 - 2.5$ and absorption column $\sim 1 - 3 \times 10^{21} \pcmsq$) the correction factor is generally in the range $1.1 - 1.3$; we used the optimistic upper value of 1.3 to correct the \chan fluxes and luminosities for each detection of each ULX in the left-hand panel.  In the right-hand panel we show the \swift data, and have chosen to display the peak flux/luminosity combination for each source (observation-level data is not present for 2SXPS sources in W22, although it is easily recoverable from 2SXPS itself).  Clearly the numbers of sources chosen from each detector as inhabiting the demarcated parameter space vary greatly.  There are only 7 CSC2.0 detections within the selection region (including one very marginal case, even after the generous correction factor).  The 5 brightest (in flux) are all already in our \xmm selection - three separate detections of 4XMM J095550.4+694045 (M82 X-1), and one each of 4XMM J022727.5+333443 (NGC 925 X-1) and 4XMM J131519.5+420301 (NGC 5055 X-1).  The two fainter sources are not in the current list, and so present new targets.  The 2SXPS data, on the other hand, presents far more potential targets than the \xmm selection - a total of 75 eULX candidates with peak flux and luminosity in the correct region of parameter space.  Ten of these were in the \xmm selection presented in Section~\ref{sec:source_select}; those missing include two of the sources presented in this paper (\pgcsrc and \cv respectively), M82 X-1 which will have been excluded by W22 for being too close the the nucleus of M82 for \swift to resolve it from any putative nuclear activity, 4XMM J151558.6+561810 (NGC 5907 ULX1) whose luminosity peaked in the HLX regime during \swift observations, and 4XMM J230457.6+122028 (NGC 7479 ULX-1) where the flux was marginally below the cut-off.  Note, however, that if we had chosen instead to use the average \swift flux and luminosity per source, we would have obtained a similar outcome to the original \xmm selection -- 15 sources (albeit the overlap with the \xmm selection was only 7 objects, with the other 8 all new potential targets).  The analysis of data from these new eULX candidates is left to future work.

The lack of \chan sources is puzzling, but may be due to a combination of several factors.  These include the narrower bandpass (although an attempt was made to correct for this in the figure); the smaller point spread function, which will be much better at removing contaminants (local diffuse emission and/or other point sources) than is possible for \xmm or {\it Swift\/}; and perhaps most importantly the removal of sources from W22 that are quality flagged, which will have removed objects at relatively high fluxes due to the presence of readout streaks and/or pile-up.  The much larger potential target list revealed by \swift is likely a consequence of two factors.  First, \swift has covered a larger sky area than either other mission (e.g. $\sim 3800 \rm~deg^2$ in 2SXPS, \citealt{evans20}, compared to $\sim 1200 \rm ~deg^2$ in 4XMM-DR10 which is marginally larger than the area covered by 4XMM-DR9 reported in \citealt{webb20}), and so has better coverage of galaxies at moderate distances ($< 30$ Mpc) for which the detection of eULX candidates is possible, even for short \swift snapshots.  Second, it takes many more observations of individual sources, on average, than the other missions.  Hence, for ULXs which are generally a variable population, it has more chances to catch a source in the correct luminosity range.  The combination of these two factors bodes well for the detection of eULXs by the {\it eROSITA\/} all-sky survey, which will combine multiple visits to the position of each ULX candidate with full sky coverage and sensitivity per visit similar to the \swift snapshots.

\subsection{Is this a good method for finding PULX candidates?}

The starting point for this work was to search for new PULX candidates, using a combination of high flux, high luminosity and existing \xmm data as the selection criteria.  This has revealed one new interesting PULX candidate, and another object requiring further observations.  There are many other potential candidates for observation revealed by {\it Swift\/}, and a couple more by {\it Chandra\/}.  However, a basic question remains - is this a good way of finding good candidates to be PULXs?  This is important to answer as it helps us to make the best use of limited observational time on X-ray facilities.  

There are a couple of alternative approaches that can be taken.  \cite{earnshaw18} suggested looking for ULXs that display large amplitude variability, based in the suggestion of \cite{tsygankov16} that high amplitude flux variability in M82 X-2 was due to the propeller effect (although it is now unclear whether this is the case, or it is instead variable due to a $\sim 60$ day super-orbital period, cf. \citealt{brightman19}).  \cite{song20} improved upon the Earnshaw work, identifying 17 good candidates including two known PULXs; however none of the new objects in that sample have yet been identified as PULXs, and the premise of looking for large amplitude variability as evidence for the propeller effect has recently been questioned by \cite{middleton23} who argue this should instead lead to increased variability in the hard spectral component of ULXs.  This is consistent with \cite{walton18}, who show that known PULXs have a dominant hard spectral component when pulsating, which reduces in strength when they are not doing so.  

Indeed, a second method for looking for PULXs is to compare the spectral and timing behaviour of other ULXs to the known PULXs; \cite{gurpide21} suggest good PULX candidates on this basis, but again no pulsations have been found from these objects yet.  However, propeller-induced variability may not be the only reason to continue to look for high amplitude variability in ULXs; \cite{khan22} show that if NS-ULXs have a higher degree of geometric beaming than BH-ULXs, then precession of the accretion discs will result in a higher fraction of variable NS-ULXs than BH-ULXs.   Therefore, high amplitude variability may remain as a possible marker for PULX candidates even without propeller effects.

In the absence of of any positive results, it is not clear whether any of these suggested strategies provide a better chance of finding detectable PULXs than the others.  It is therefore prudent to continue to attempt them all if we wish to maximise our chances of discovering more PULXs, and thereby learn more about the physics of super-Eddington accretion in the regime of high magnetic field compact objects.

\section{Conclusions}

We started by posing a question - can we use the fact that most of the small number of PULXs detected to date have observed luminosities that peak in excess $10^{40} \ergs$ (i.e. in the eULX regime) to find more PULX candidates?  To answer this, we filtered the ULX catalogue of \cite{walton22} to select objects with \xmm data showing peak luminosities in this range, concurrent with fluxes above $5 \times 10^{-13} \ergcms$ (with the latter used to select objects where we have a good chance of accumulating $> 10^4$ counts in an orbit, sufficient to detect pulsations).  Most of the 15 objects selected are well-studied ULXs; however we uncovered three ULXs that have hitherto been largely neglected for study.  The analysis of available archival data for these objects is the subject of this paper.

We found \cv to be a soft and variable X-ray source.  It lies at low Galactic latitude and is coincident with a bright optical/IR source with measured parallax and $\sim 2$ day periodic variability, indicating it is very likely a foreground contaminant.  However, a full diagnosis of the nature of this object requires more observations.  The presence of a Galactic contaminant highlights a weakness in the filtering of W22 for such objects.  However, this could readily be corrected by the use of {\it Gaia\/} data and/or the use of statistical/multiwavelength identification techniques to better identify foreground objects in future catalogues, as has already been done in the work of \cite{bernadich22} and \cite{tranin23}.  \pgcsrc on the other hand remains an excellent eULX candidate, but suffers from minimal archival data being available at present, so also requires further observational study to improve our understanding of it.  The final object, \ngcsrc has by far the best current archival data.  It shows moderate X-ray variability on timescales of minutes to years, an X-ray spectrum that is hard and can be modelled similarly to other ULXs, and a local environment in the spiral arm of its host galaxy that appears similar to other ULX environments (including potential counterparts, which indicate an X-ray-to-optical flux ratio that is too high for a background AGN).  We are able to perform accelerated pulsation searches on the \xmm pn data for this object, but we do not find any pulsations, with the best limits of a pulse fraction $\lesssim 20\%$ for a pulsation at $\sim 1$ Hz.  So, \ngcsrc is not at this time a confirmed PULX; it does however remain a plausible candidate given its hard X-ray spectrum, moderate limits on the pulsed fraction, and the fact that pulsations are transient characteristics of PULXs.  Again, further observations with the current fleet of X-ray observatories are required to better understand the nature of this object.  

Finally, we note that the selection of high luminosity, high flux ULXs provides an interesting sample of objects that could be the subject of further study with many future X-ray missions, both imminent and proposed (e.g. {\it XRISM\/}, {\it HEX-P\/}, {\it New Athena\/} etc.), that could provide new opportunities to investigate whether they host PULXs.  We look forward to the excellent science they will enable.

\section*{Acknowledgements}

We thank the anonymous referee for their constructive comments that have helped improve this paper.  TPR \& SS were supported by the Science and Technology Facilities Council (STFC) as part of consolidated grant ST/T000244/1.  ADAM thanks STFC for their support via an STFC studentship (ST/T506047/1).  This research has made use of data obtained with {\it XMM-Newton\/}, an ESA science mission with instruments and contributions directly funded by ESA Member States and NASA.  It has made use of data obtained from the \chan Data Archive and the \chan Source Catalog, and software provided by the \chan X-ray Center (CXC) in the application package {\sc ciao}.  This work made use of data supplied by the UK \swift Science Data Centre at the University of Leicester.  It uses observations made with the NASA/ESA {\it Hubble Space Telescope}, and obtained from the Hubble Legacy Archive, which is a collaboration between the Space Telescope Science Institute (STScI/NASA), the Space Telescope European Coordinating Facility (ST-ECF/ESAC/ESA) and the Canadian Astronomy Data Centre (CADC/NRC/CSA).  This paper includes data collected with the {\it TESS} mission, obtained from the MAST data archive at the STScI. Funding for the {\it TESS} mission is provided by the NASA Explorer Program. STScI is operated by the Association of Universities for Research in Astronomy, Inc., under NASA contract NAS 5-26555.  The Pan-STARRS1 Surveys (PS1) and the PS1 public science archive have been made possible through contributions by the Institute for Astronomy, the University of Hawaii, the Pan-STARRS Project Office, the Max-Planck Society and its participating institutes, the Max Planck Institute for Astronomy, Heidelberg and the Max Planck Institute for Extraterrestrial Physics, Garching, The Johns Hopkins University, Durham University, the University of Edinburgh, the Queen's University Belfast, the Harvard-Smithsonian Center for Astrophysics, the Las Cumbres Observatory Global Telescope Network Incorporated, the National Central University of Taiwan, the Space Telescope Science Institute, the National Aeronautics and Space Administration under Grant No. NNX08AR22G issued through the Planetary Science Division of the NASA Science Mission Directorate, the National Science Foundation Grant No. AST-1238877, the University of Maryland, Eotvos Lorand University (ELTE), the Los Alamos National Laboratory, and the Gordon and Betty Moore Foundation.  This work has made use of data from the European Space Agency (ESA) mission {\it Gaia} (\url{https://www.cosmos.esa.int/gaia}), processed by the {\it Gaia} Data Processing and Analysis Consortium (DPAC, \url{https://www.cosmos.esa.int/web/gaia/dpac/consortium}).  Funding for the DPAC has been provided by national institutions, in particular the institutions participating in the {\it Gaia} Multilateral Agreement.  This research has made use of the NASA/IPAC Extragalactic Database (NED) which is operated by the Jet Propulsion Laboratory, California Institute of Technology, under contract with the National Aeronautics and Space Administration.  It has also made use of NASA's Astrophysics Data System.


\section*{Data Availability}

The data used in this paper is all available from public archives, via searches for the observation identifiers in Table~\ref{tab:data} and/or the text, or via searches at the positions of the three sources studied..  \xmm data was obtained from the \xmm Science Archive (XSA) (\url{https://www.cosmos.esa.int/web/xmm-newton/xsa}), and \chan data was obtained from the \chan Data Archive (\url{https://cxc.cfa.harvard.edu/cda/}).  The \swift data was all obtained from 2SXPS (\url{https://www.swift.ac.uk/2SXPS/}).  Pan-STARRS imaging data was obtained from \url{http://ps1images.stsci.edu/cgi-bin/ps1cutouts}, and \hst data was obtained from MAST (\url{https://archive.stsci.edu}).  The TESS data used in the analysis of this work is available on the MAST webpage (\url{https://mast.stsci.edu/portal/Mashup/Clients/Mast/Portal.html}).  {\it Gaia} data was obtained via the VizieR interface for astronomical catalogues (\url{https://vizier.cds.unistra.fr}).
 



\bibliographystyle{mnras}
\bibliography{example} 








\bsp	
\label{lastpage}
\end{document}